\documentclass[preprint,longtitle,12pt,twocolumn,3p]{elsarticle}
\usepackage{amssymb}
\journal{Astroparticle Physics}
\usepackage{lineno}
\usepackage[hyperindex]{hyperref}
\usepackage{breqn}
\usepackage{subcaption}
\usepackage{lineno}
\usepackage{makecell}
\usepackage{relsize}
\input{se.def}
\newcommand{\mus}{\mu \mathrm{s}}
\input{my_ds_macro.def}

\begin{document}
\begin{frontmatter}
    \title{A study of events with photoelectric emission  in the DarkSide-50 liquid argon Time Projection Chamber}
    \newcommand{\Alberta}{Department of Physics, University of Alberta, Edmonton, AB T6G 2R3, Canada}
\newcommand{\APC}{APC, Universit\'e de Paris, CNRS, Astroparticule et Cosmologie, Paris F-75013, France}
\newcommand{\AQLNGS}{INFN Laboratori Nazionali del Gran Sasso, Assergi (AQ) 67100, Italy}
\newcommand{\AQGSSI}{Gran Sasso Science Institute, L'Aquila 67100, Italy}
\newcommand{\AQUni}{Università degli Studi dell'Aquila, L'Aquila 67100, Italy}
\newcommand{\AUM}{InstitutodeF\'õsica,Universidad Nacional Auto\'nomade M\'exico(UNAM), M\'exico 01000, Mexico}
\newcommand{\AstroCeNT}{AstroCeNT, Nicolaus Copernicus Astronomical Center, 00-614 Warsaw, Poland}
\newcommand{\Augustana}{Physics Department, Augustana University, Sioux Falls, SD 57197, USA}
\newcommand{\Belgorod}{Radiation Physics Laboratory, Belgorod National Research University, Belgorod 308007, Russia}
\newcommand{\BHSU}{School of Natural Sciences, Black Hills State University, Spearfish, SD 57799, USA}
\newcommand{\BINP}{Budker Institute of Nuclear Physics, Novosibirsk 630090, Russia}
\newcommand{\BNL}{Brookhaven National Laboratory, Upton, NY 11973, USA}
\newcommand{\BOINFN}{INFN Bologna, Bologna 40126, Italy}
\newcommand{\BOUniPHY}{Physics Department, Universit\`a degli Studi di Bologna, Bologna 40126, Italy}
\newcommand{\CAUniCHE}{Department of Mechanical, Chemical, and Materials Engineering, Universit\`a degli Studi, Cagliari 09042, Italy}
\newcommand{\CAUniPHY}{Physics Department, Universit\`a degli Studi di Cagliari, Cagliari 09042, Italy}
\newcommand{\CAINFN}{INFN Cagliari, Cagliari 09042, Italy}
\newcommand{\Carleton}{Department of Physics, Carleton University, Ottawa, ON K1S 5B6, Canada}
\newcommand{\Campinas}{Physics Institute, Universidade Estadual de Campinas, Campinas 13083, Brazil}
\newcommand{\CentroFermi}{Museo della fisica e Centro studi e Ricerche Enrico Fermi, Roma 00184, Italy}
\newcommand{\CIEMAT}{CIEMAT, Centro de Investigaciones Energ\'eticas, Medioambientales y Tecnol\'ogicas, Madrid 28040, Spain}
\newcommand{\Cluj}{National Institute for R\&D of Isotopic and Molecular Technologies, Cluj-Napoca, 400293, Romania}
\newcommand{\CPPM}{Centre de Physique des Particules de Marseille, Aix Marseille Univ, CNRS/IN2P3, CPPM, Marseille, France}
\newcommand{\CTLNS}{INFN Laboratori Nazionali del Sud, Catania 95123, Italy}
\newcommand{\ENUniCEE}{Engineering and Architecture Faculty, Universit\`a di Enna Kore, Enna 94100, Italy}
\newcommand{\ETHZ}{Institute for Particle Physics, ETH Z\"urich, Z\"urich 8093, Switzerland}
\newcommand{\FNAL}{Fermi National Accelerator Laboratory, Batavia, IL 60510, USA}
\newcommand{\FortLewis}{Department of Physics and Engineering, Fort Lewis College, Durango, CO 81301, USA}
\newcommand{\GEUni}{Physics Department, Universit\`a degli Studi di Genova, Genova 16146, Italy}
\newcommand{\GEINFN}{INFN Genova, Genova 16146, Italy}
\newcommand{\GlenEllyn}{Glen Ellyn, Illinois 60137, USA}
\newcommand{\Hawaii}{Department of Physics and Astronomy, University of Hawai'i, Honolulu, HI 96822, USA}
\newcommand{\Houston}{Department of Physics, University of Houston, Houston, TX 77204, USA}
\newcommand{\IHEP}{Institute of High Energy Physics, Beijing 100049, China}
\newcommand{\IPNO}{Institut de Physique Nucl\`eaire dÕOrsay, 91406, Orsay, France}
\newcommand{\INSTM}{Interuniversity Consortium for Science and Technology of Materials, Firenze 50121, Italy}
\newcommand{\IPHC}{IPHC, Universit\'e de Strasbourg, CNRS/IN2P3, Strasbourg 67037, France}
\newcommand{\JINR}{Joint Institute for Nuclear Research, Dubna 141980, Russia}
\newcommand{\Krakow}{M. Smoluchowski Institute of Physics, Jagiellonian University, 30-348 Krakow, Poland}
\newcommand{\Kurchatov}{National Research Centre Kurchatov Institute, Moscow 123182, Russia}
\newcommand{\Laurentian}{Department of Physics and Astronomy, Laurentian University, Sudbury, ON P3E 2C6, Canada}
\newcommand{\LNFINFN}{INFN Laboratori Nazionali di Frascati, Frascati 00044, Italy}
\newcommand{\Lodz}{Institute of Applied Radiation Chemistry, Lodz University of Technology, 93-590 Lodz, Poland}
\newcommand{\LPNHE}{LPNHE, CNRS/IN2P3, Sorbonne Universit\'e, Universit\'e Paris Diderot, Paris 75252, France}
\newcommand{\Manchester}{The University of Manchester, Manchester M13 9PL, United Kingdom}
\newcommand{\MEPhI}{National Research Nuclear University MEPhI, Moscow 115409, Russia}
\newcommand{\MIBIINFN}{INFN Milano Bicocca, Milano 20126, Italy}
\newcommand{\MIINFN}{INFN Milano, Milano 20133, Italy}
\newcommand{\MIPoliICA}{Civil and Environmental Engineering Department, Politecnico di Milano, Milano 20133, Italy}
\newcommand{\MIPoliCHE}{Chemistry, Materials and Chemical Engineering Department ``G.~Natta", Politecnico di Milano, Milano 20133, Italy}
\newcommand{\MIPoliEIB}{Electronics, Information, and Bioengineering Department, Politecnico di Milano, Milano 20133, Italy}
\newcommand{\MIPoliENE}{Energy Department, Politecnico di Milano, Milano 20133, Italy}
\newcommand{\MIUni}{Physics Department, Universit\`a degli Studi di Milano, Milano 20133, Italy}
\newcommand{\MSU}{Skobeltsyn Institute of Nuclear Physics, Lomonosov Moscow State University, Moscow 119234, Russia}
\newcommand{\NAINFN}{INFN Napoli, Napoli 80126, Italy}
\newcommand{\NAUniPHY}{Physics Department, Universit\`a degli Studi ``Federico II'' di Napoli, Napoli 80126, Italy}
\newcommand{\NAUniCHE}{Chemical, Materials, and Industrial Production Engineering Department, Universit\`a degli Studi ``Federico II'' di Napoli, Napoli 80126, Italy}
\newcommand{\NSU}{Novosibirsk State University, Novosibirsk 630090, Russia}
\newcommand{\OACINAF}{INAF Osservatorio Astronomico di Capodimonte, 80131 Napoli, Italy}
\newcommand{\Petersburg}{Saint Petersburg Nuclear Physics Institute, Gatchina 188350, Russia}
\newcommand{\PGUniCBB}{Chemistry, Biology and Biotechnology Department, Universit\`a degli Studi di Perugia, Perugia 06123, Italy}
\newcommand{\PGINFN}{INFN Perugia, Perugia 06123, Italy}
\newcommand{\PIINFN}{INFN Pisa, Pisa 56127, Italy}
\newcommand{\PIUniPHY}{Physics Department, Universit\`a degli Studi di Pisa, Pisa 56127, Italy}
\newcommand{\PNNL}{Pacific Northwest National Laboratory, Richland, WA 99352, USA}
\newcommand{\Princeton}{Physics Department, Princeton University, Princeton, NJ 08544, USA}
\newcommand{\Queens}{Department of Physics, Engineering Physics and Astronomy, QueenÕs University, Kingston, ON K7L 3N6, Canada}
\newcommand{\RHUL}{Department of Physics, Royal Holloway University of London, Egham TW20 0EX, UK}
\newcommand{\RMTreINFN}{INFN Roma Tre, Roma 00146, Italy}
\newcommand{\RMTreUni}{Mathematics and Physics Department, Universit\`a degli Studi Roma Tre, Roma 00146, Italy}
\newcommand{\RMUnoINFN}{INFN Sezione di Roma, Roma 00185, Italy}
\newcommand{\RMUnoUni}{Physics Department, Sapienza Universit\`a di Roma, Roma 00185, Italy}
\newcommand{\SAINFN}{INFN Salerno, Salerno 84084, Italy}
\newcommand{\SNOLabaddress}{SNOLAB, Lively, ON P3Y 1N2, Canada}
\newcommand{\SNOLAB}{SNOLAB, Lively, ON P3Y 1N2, Canada}
\newcommand{\SSUniCHP}{Chemistry and Pharmacy Department, Universit\`a degli Studi di Sassari, Sassari 07100, Italy}
\newcommand{\Sussex}{Physics and Astronomy, University of Sussex, Brighton BN1 9QH, UK}
\newcommand{\Temple}{Physics Department, Temple University, Philadelphia, PA 19122, USA}
\newcommand{\TNFBK}{Fondazione Bruno Kessler, Povo 38123, Italy}
\newcommand{\TNTIFPA}{Trento Institute for Fundamental Physics and Applications, Povo 38123, Italy}
\newcommand{\TNUni}{Physics Department, Universit\`a degli Studi di Trento, Povo 38123, Italy}
\newcommand{\TOINFN}{INFN Torino, Torino 10125, Italy}
\newcommand{\TOPoli}{Department of Electronics and Communications, Politecnico di Torino, Torino 10129, Italy}
\newcommand{\TOUni}{Physics Department, Universit\`a degli Studi di Torino, Torino 10125, Italy}
\newcommand{\TRIUMFaddress}{TRIUMF, 4004 Wesbrook Mall, Vancouver, British Columbia V6T2A3, Canada}
\newcommand{\TUM}{Physik Department, Technische Universit\"at M\"unchen, Munich 80333, Germany}
\newcommand{\UB}{Universiatat de Barcelona, Barcelona E-08028, Catalonia, Spain} 
\newcommand{\UCDavis}{Department of Physics, University of California, Davis, CA 95616, USA}
\newcommand{\UCLA}{Physics and Astronomy Department, University of California, Los Angeles, CA 90095, USA}
\newcommand{\UMass}{Amherst Center for Fundamental Interactions and Physics Department, University of Massachusetts, Amherst, MA 01003, USA}
\newcommand{\UOC}{Department of Chemistry, University of Crete, P.O. Box 2208, 71003 Heraklion, Crete, Greece}
\newcommand{\USP}{Instituto de F\'isica, Universidade de S\~ao Paulo, S\~ao Paulo 05508-090, Brazil}
\newcommand{\VTech}{Virginia Tech, Blacksburg, VA 24061, USA}
    \author[1]{P.~Agnes}
\author[4]{I.~F.~M.~Albuquerque}
\author[5]{T.~Alexander}
\author[8]{A.~K.~Alton}
\author[4]{M.~Ave}
\author[5]{H.~O.~Back}
\author[17,18]{G.~Batignani}
\author[66]{K.~Biery}
\author[40]{V.~Bocci}
\author[21]{W.~M.~Bonivento}
% ...

\address[1]{\Houston}
\address[4]{\USP}
\address[5]{\PNNL}
\address[8]{\Augustana}
% ...
\address[17]{\PIINFN}
\address[18]{\PIUniPHY}
\address[21]{\CAINFN}
% ...
\address[40]{\RMUnoINFN}
% ...
\address[66]{\FNAL}
% ...
\fntext[fn1]{Deceased.}
    \date{\today}
    \begin{abstract}
Finding unequivocal evidence of dark matter interactions in a particle detector is a major objective of physics research. Liquid argon time projection chambers offer a path to probe   Weakly Interacting Massive Particles scattering  cross sections  on nucleus   down to the so-called neutrino floor, in a mass range from few GeV's to hundredths of TeV's. Based on the successful operation of the DarkSide-50 detector at LNGS, a new and more sensitive experiment, DarkSide-20k, has been designed and is now under construction. A thorough understanding of the DarkSide-50 detector response and, therefore,  of all kind of observed events, 
is essential for an optimal design of the new experiment. In this paper, we report on a particular set of events, which were not used for  dark matter searches. Namely, standard two-pulse scintillation-ionization signals accompanied by a small amplitude third pulse, originating from single or  few  electrons, in a time window of less than a maximum drift time.  
We compare our findings to those of a recent paper of the LUX Collaboration (D.S.Akerib et al. Phys.Rev.D 102, 092004). Indeed, both experiments observe events related to  photoionization of the cathode.
From the measured rate of these events, we estimate for the first time the quantum efficiency of the tetraphenyl butadiene deposited on the \DSf\ cathode  at wavelengths around 128~nm, in liquid argon.
Also, both experiments observe  events likely related to photoionization of impurities in the liquid. The probability of photoelectron emission per unit length turns out to be  one order of magnitude smaller in \DSf\ than in LUX. This result,   together with the much larger measured electron lifetime, coherently hints toward  a lower concentration of contaminants in \DSf\ than in LUX.

    \end{abstract}
\end{frontmatter}

\section{Introduction}
\label{sec:intro}

Direct detection of Weakly Interacting Massive Particle Dark Matter (WIMP DM) 
is one of the most active areas of astroparticle physics.
The Liquid Argon (LAr) Time Projection Chamber (TPC) 
technology offers a path to reach sensitivities to WIMP-nucleus scattering cross-sections down to the so-called neutrino floor~\cite{NF}, for both high and low WIMP masses.

Based on the successful operation of the \DSf\ (\DSfs) detector \cite{Agnes:2018fg,Agnes:2018ft}, a larger  and more sensitive experiment, \DSk\ (\DSks) \cite{Aalseth:2018gq}, is  now under construction. A deep understanding of the \DSfs\ detector response is one of  key ingredients for an 
optimal design of \DSks. Therefore, beyond studying events related 
to dark matter searches, it is very 
 important to scrutinize all event types  in the detector, since they may provide hints for detector optimization.

A typical interaction in the active volume of the TPC yields a prompt scintillation signal, S1, and one or more clouds of ionization electrons, depending on the single- or multi-scatter nature of the interaction. In the \DSfs\ LAr TPC, ionization electrons drift upwards under a uniform electric field and are, under the application of   two other fields, extracted into the  gas pocket  and induce one or more  electroluminescence signals, S2. 
As discussed in~\cite{Agnes:2018hvf}, S1 and S2 signals have different pulse shapes. The S1 signal  rises in  few ns and falls as a double exponential, with $\tau_1=(6\pm 1)$~ns and  $\tau_2=(1.5\pm 0.1)~\mu$s, and an  amplitude ratio of the two exponentials of  $\sim3$ for nuclear recoils and $\sim0.3$ for electron recoils \cite{Boulay:2006hu,PhysRevB.27.5279}. This difference in amplitude ratios leads to a very effective Pulse Shape Discrimination (PSD) between electron and nuclear recoils.
The S2 signal has a different pulse shape, i.e. a $\sim 1~\mu$s rise-time and a $\sim 3~\mu$s fall-time. 
The detection of  both  S1 and S2 pulses  allows three-dimensional reconstruction of the interaction point and, therefore,  background rejection by detection of  multiple interactions 
and  volume fiducialization. 
In \DSfs\ the typical pulse charge ratio S2 over S1, 
as discussed in \refsec{detector}, is between $10$ and  $30$.
Therefore, low energy interactions may  yield  only  S2 signals  above detection threshold. These single-pulse events   were exploited to extend dark matter searches to  lower masses \cite{Agnes:2018fg}.

In addition to these {\em standard}  events, other event types were  observed in the \DSfs\ detector.
In this paper, we discuss {\em prompt} emission events, namely
 events with an additional small amplitude S2 pulse, referred to as Single Electron Candidate (SEC) in the following, occurring in the same $440~\mus$ data acquisition window of standard events. 
 We classify these events into two different categories:  {\em echo  events}, discussed in \refsec{corre}, when the SEC   has   a definite temporal relationship  with the preceding S1 or  S2,  and {\em bulk  events}, discussed in \refsec{liquid}, when the SEC  does not have a definite temporal relationship with the preceding S1 or  S2, but it is consistent with being  due to one single  electron.
Therefore, both these  event types  have   features
that  clearly distinguish them from the trivial multi-scatter photon background 
interactions. 

Events with  single electron signals occurring outside of the acquisition window of  a previous standard event, i.e. due to a {\em delayed} emissions,  were also observed in \DSfs~\cite{Agnes:2018fg}  and will be further  discussed and analyzed in an upcoming \DSfs\ publication. 

Based on the study detailed in this paper, we also provide an 
  interpretation of  observed event types. 
%

Similar kinds of events as those discussed in this paper were also observed and studied with xenon detectors. The most comprehensive study was performed  by the LUX Collaboration \cite{PhysRevD.102.092004} and we will use it for comparison with our results.
Other previous papers reporting similar event types can be found in Refs.~\cite{Edwards:2007nj, Aprile:2013blg, Akimov_2016}.

\section{The \DSf\ detector}
\label{sec:detector}

The \DSfs\ LAr TPC detects light from both \SOne\ and \STwo\ using 
\DSfPMTsNumber\ \DSfPMTSize\ photo-multipliers (\PMTs) arranged in two arrays of 19 \PMTs\ each, at both ends of 
the \DSfActiveMass\ cylindrical active target of low-radioactivity underground argon (\UAr)~\cite{Agnes:2018ep,AcostaKane:2008im,Xu:2015do}. The \PMTs\ are submerged in liquid argon and view the active volume through fused silica windows. These are coated on both faces with transparent conductive indium tin oxide (ITO) films \SI{15}{\nano\meter} thick. 
The inner window faces define the grounded anode (top) and HV cathode (bottom) of the TPC, while the outer faces are kept at the average photocathode potential of each 19-PMT array.
The cylindrical side wall is made of 2.54 cm-thick polytetrafluoroethylene (PTFE) reflector sintered using a special annealing cycle to increase its reflectivity. The PTFE reflector and the fused silica windows 
are coated with tetraphenyl butadiene (TPB) wavelength shifter, which absorbs the \SI{128}{\nano\meter} LAr scintillation photons 
and re-emits visible photons with a peak wavelength of \SI{420}{\nano\meter}. The specific thickness of the TPB coating on the windows varies between $(230 \pm 10)~\mu \mathrm{g}/\mathrm{cm}^2$ at the center and $(190 \pm 15)~\mu \mathrm{g}/\mathrm{cm}^2$ at the edge of the active volume, corresponding to few $\mu$m thickness. The thickness of the TPB on the cylindrical wall is $(165 \pm 20)~\mu \mathrm{g}/\mathrm{cm}^2$ at half-height and $(224 \pm 27)~\mu \mathrm{g}/\mathrm{cm}^2$ at the top and bottom.
The electric fields needed for drifting and extracting electrons consists of the ITO-coated cathode and anode planes, a field cage comprising a stack of copper rings behind the PTFE reflector held at graded potentials, and a grid that separates the drift and electron extraction regions. The grid, placed 5 mm below the liquid surface, is a hexagonal mesh photo-etched from a 50~$\mu$m-thick stainless steel foil and has an optical transparency of 95\% at normal incidence. 

The data reported in this paper were acquired between July 2015 and October 2017, using a TPC drift field of \DSfDriftField, an extraction field of \DSfExtractionField, and an electroluminescence field of \DSfMultiplicationField.  At this extraction field, the efficiency for extracting ionization electrons into the gas layer is estimated to be close to 100\% ~\cite{Bondar_2009,Gushchin:1982b}. 
The  electron  drift time,   $t_{\mathrm{drift}}=\Delta t_{\mathrm{S2}-\mathrm{S1}}$, has a maximum value at $t_{\mathrm{drift}}^{\mathrm{max}}$=376~$\mus$, for interactions  located right above  the cathode. The electron drift speed  is $(0.93 \pm 0.01)~\mathrm{mm}/\mus$~\cite{Agnes_2017}. 

A hardware  trigger in \DSfs\ occurs when  two or more PMT signals exceed a threshold of 0.6 Photo-Electrons (PE) within a \DSfTriggerWindow\ window.
Waveform data are recorded from all 38 PMTs for $440~\mus$ starting $\sim10~\mus$ before the trigger. 
Subsequent triggers are inhibited for $810~\mus$.
Software pulse-finding algorithms are then applied to the digitized data, including the pre-trigger data. The software classifies the pulses into two categories (\SOne\ or \STwo) based on the fraction of light detected within the first \WindowFNine\ (\FNine).
The efficiency of the software pulse-finding algorithm is essentially \DSfPulseFindingEfficiency\ for \STwo\ signals larger than \DSfPulseFindingEfficiencyThreshold~\cite{Agnes:2015gu}.
The pulse finder uses an integration window of $30~\mus$, which is long enough to include the entire \STwo\ signal.

The argon is  purified continuously by recirculating it in gaseous form through a heated getter (SAES Monotorr PS4-MT50-R-2), which reduces contaminants such as O$_2$ and N$_2$ to sub-ppb levels, and through a cold charcoal radon trap.
The measured electron  lifetime was larger than $\sim$\SI{8}{ms} during the whole data-taking, corresponding to   $\sim$\SI{35}{ppt} O$_{2}$-equivalent contamination \cite{Acciarri_2010}. 

\section{Event selection}
\label{sec:categorization}
We select three-pulse events,
with an S1  
followed by two S2.  The S1 pulse provides  the event trigger.
One of the S2 pulses, the SEC, is required to have a charge 
smaller than 
200~PE.
 
We  require  the event trigger to occur at least $400~\mus$ after the end of the inhibit window of the previous trigger, namely at least 1.21 ms after the previous trigger. This removes events which triggered on an S2, with the corresponding S1   occurring during the inhibit window~\cite{Agnes:2018ep}.

The S2 light yield   drops  by about   60\%  from the center to the sides of the detector~\cite{Agnes:2017grb}. To avoid    efficiency corrections, we only select events with the maximum of the SEC signal  in the central top PMT. The corresponding effective surface cathode area is a circle of about 9 cm diameter.
Moreover, to simplify efficiency calculations, we select events with the maximum of the S2 signal in one of the 19 top PMTs.

The trigger, as shown in Ref.\cite{Agnes:2018fg}, is fully efficient for pulses above 30 PE, and, since all  events studied here are triggered by S1, the trigger inefficiency is completely negligible.

The  \FNine\ variable 
is also used to distinguish between  electron and nuclear recoils. Indeed, for electron recoils its value  clusters around 0.3 while for nuclear recoils  around 0.7.
For the rest of this paper, we restrict our selection to electron recoil events,
by requiring \FNine$<0.5$. Moreover, to limit saturation effects and pulse overlaps we require S2$<$50,000 PE and 100 PE $<$S1$<$1500 PE.

We classify the selected events into two groups, according to the time sequence of the three pulses:
S1-S2-SEC, with the SEC occurring after the S2 pulse, and 
S1-SEC-S2, with the SEC occurring between S1 and S2.   

To further strengthen the correct identification of the pulse sequence, we require the ratio of S2 to S1  to be larger than 10, as expected when the two pulses come from the same electron recoil interaction. Indeed, in \DSfs,   the typical S2 to S1 charge ratio for electron recoils is between 10 and 30.  

\section{Echo events}
\label{sec:corre}
\begin{figure}
\begin{center}
\includegraphics[width=\columnwidth]{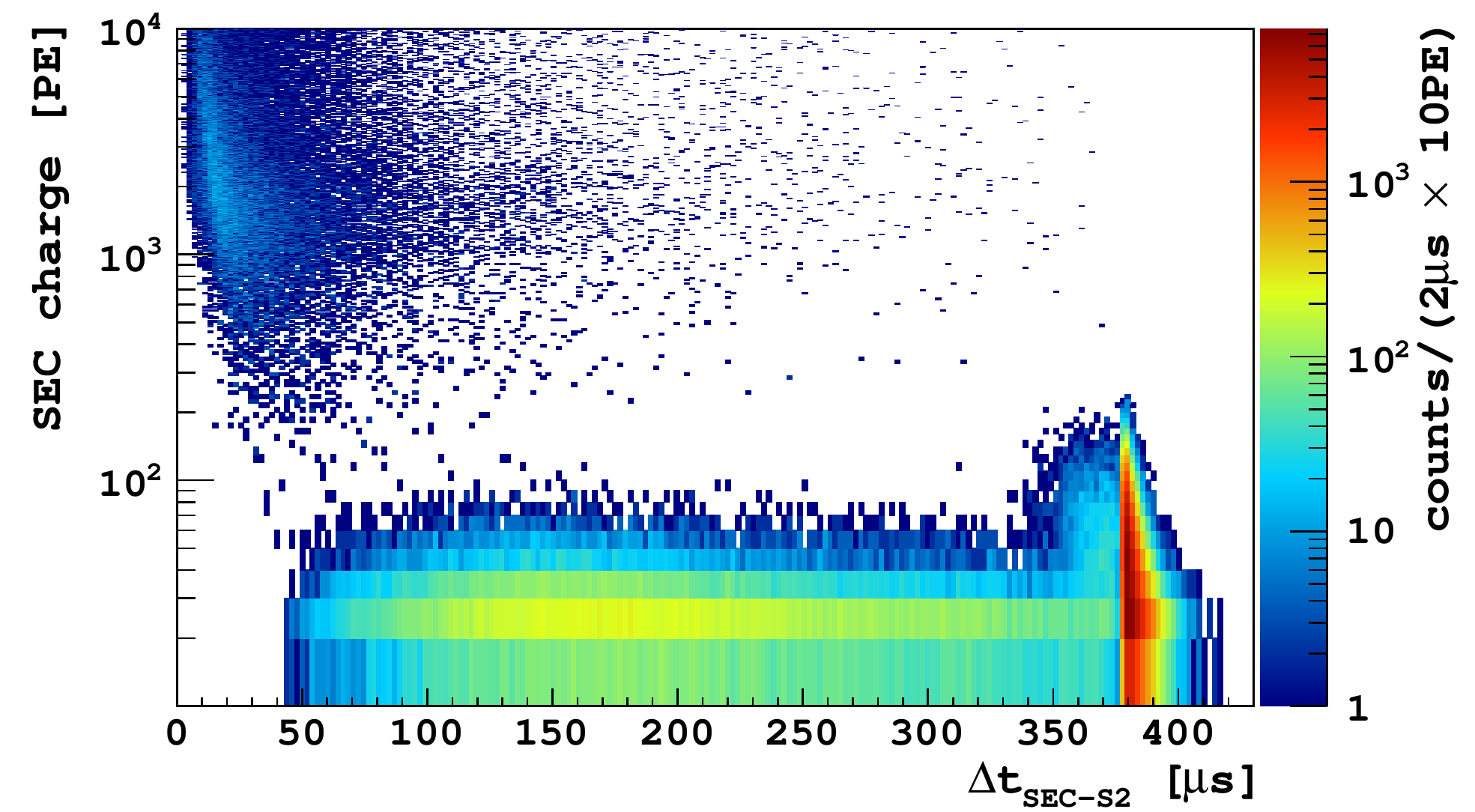}
\caption{SEC charge vs. time difference  between the SEC and the preceding S2 pulse, $\Delta t_{\mathrm{SEC}-\mathrm{S2}}$.The set of events at small values of $\Delta t_{\mathrm{SEC}-\mathrm{S2}}$ and large values of charge is related to double-scatter $\gamma$-ray interactions.}
\label{fig:SEC-S2}
\end{center}
\end{figure}

For S1-S2-SEC events, \reffig{SEC-S2} shows the charge of the SEC pulse vs. the time difference, $\Delta t_{\mathrm{SEC}-\mathrm{S2}}$, between the SEC and the preceding S2. We observe three main features in the plot, corresponding to three sets  of events, which will be detailed in the following sections. 
\subsection{S2-echo events}
\label{sec:S2echo}
\begin{figure}
\begin{center}
\includegraphics[width=\columnwidth]{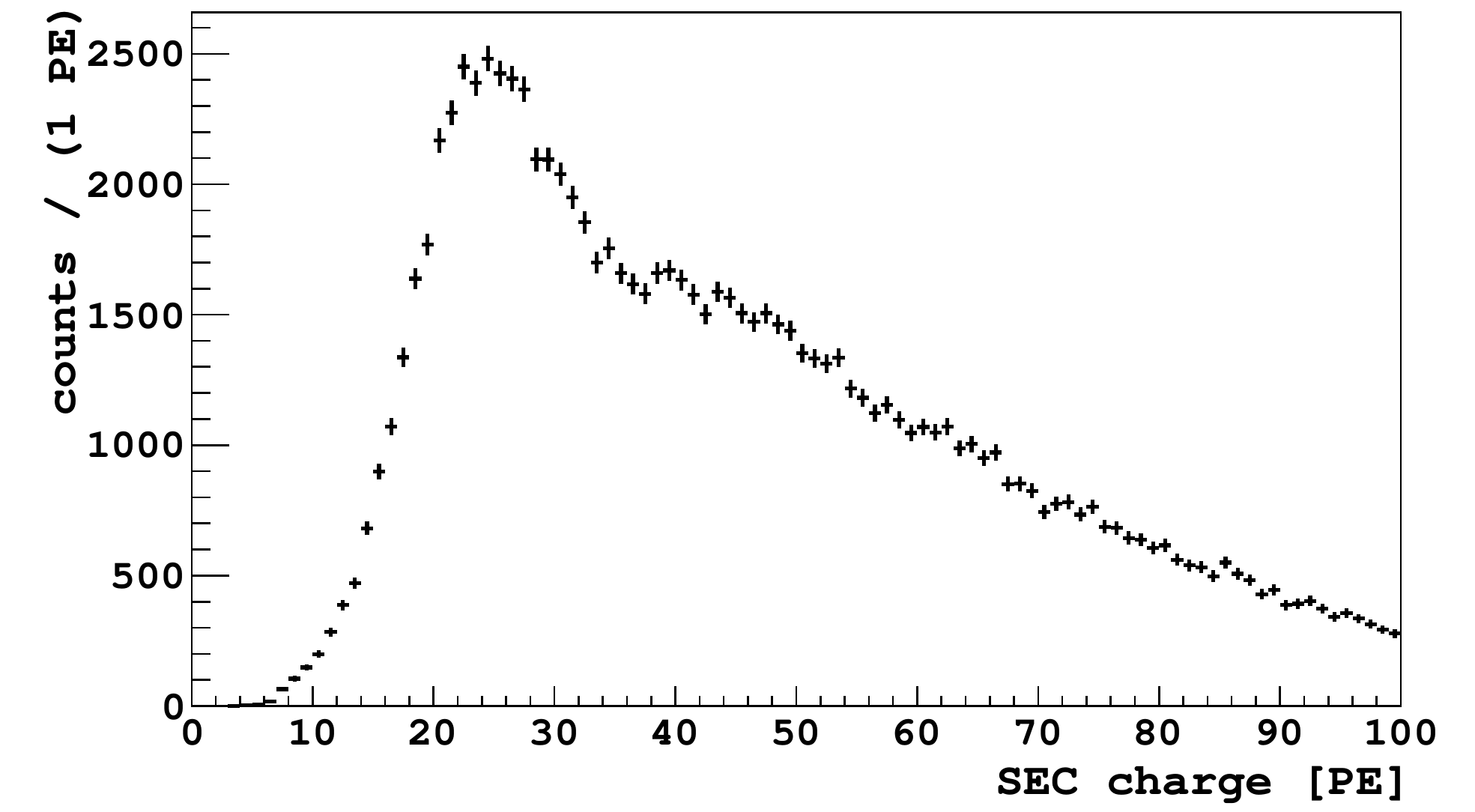}
 \caption{SEC charge spectrum for S2-echo events. }
 \label{fig:sec_charge_S2echo}
\end{center}
\end{figure}

One set of events in \reffig{SEC-S2} is clustered around   $\Delta t_{\mathrm{SEC}-\mathrm{S2}} \sim 380~\mu\mathrm{s}$, corresponding to about the  maximum TPC drift time, and  SEC charges 
extending up to a few hundred PEs. 
It seems plausible  that these events are due to  S2 photons extracting   electrons  from the cathode. The electrons then  drift under the electric field  through the whole TPC length.  Moreover, the S2 pulses are quite large signals and, sometimes, more than one photon  is able to induce  electron emission from the cathode. We call these events {\em S2-echo} events.
\reffiginitpar{sec_charge_S2echo} shows the SEC charge spectrum for these events. The peak corresponding to one electron is visible and its corresponding SEC charge is in agreement with the observation of a  previous \DSfs\ paper~\cite{Agnes:2018fg} of $\sim 23$ PE. Also, the distribution shows a tail extending to few  electron signal.

\begin{figure}
\begin{center}
\includegraphics[width=\columnwidth]{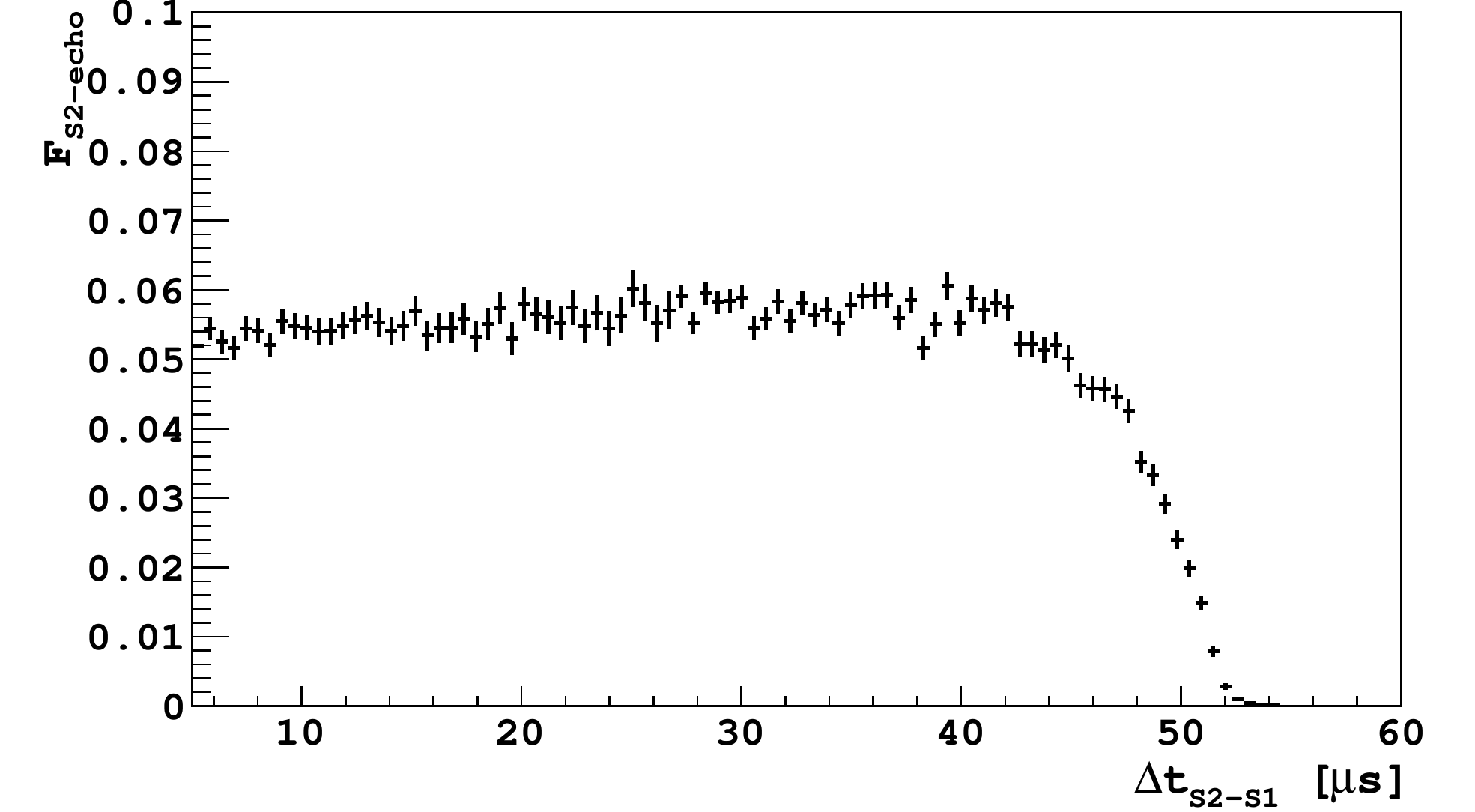}
 \caption{Fraction of  events with  an S2-echo, as a function of the drift time, $\Delta t_{\mathrm{S2}-\mathrm{S1}}$.
 }
\label{fig:S2/S1_echo_vs_z}
\end{center}
\end{figure}

The number of  S2-echo events recorded on disk is affected by the  data acquisition time window of $430~\mus$ after the trigger. This time window is smaller than $2t_{\mathrm{drift}}^{\mathrm{max}}$, the time that would be required for recording all S2-echo events. 
Indeed, when requiring three-pulse events, the \DSfs\ data acquisition only record S2-echo events originating from  interactions  in the top section of the TPC, with S1-S2 drift times, $\Delta t_{\mathrm{S2}-\mathrm{S1}}$, smaller than $ 430~\mus - t_{\mathrm{drift}}^{\mathrm{max}} \sim 50~\mus$. \reffiginitpar{S2/S1_echo_vs_z} shows the fraction of    events containing an S2-echo, as a function of the drift time, i.e.: 
\begin{dmath}
{F_{\mathrm{S2-echo}}}(t_{\mathrm{drift}}) = N_{\mathrm{S2-echo}}(t_{\mathrm{drift}}) / {N}_{\mathrm{S2}}(t_{\mathrm{drift}}).  
\label{eq:defS2echo} 
\end{dmath}
The drift time, $\Delta t_{\mathrm{S2}-\mathrm{S1}}$, is of course a measurement of the  depth of the interaction,   $z$, with $z=0$ corresponding to $\Delta t_{\mathrm{S2}-\mathrm{S1}}=0$, i.e. the gas-liquid interface.

\begin{figure}
\begin{center}
\includegraphics[width=\columnwidth]{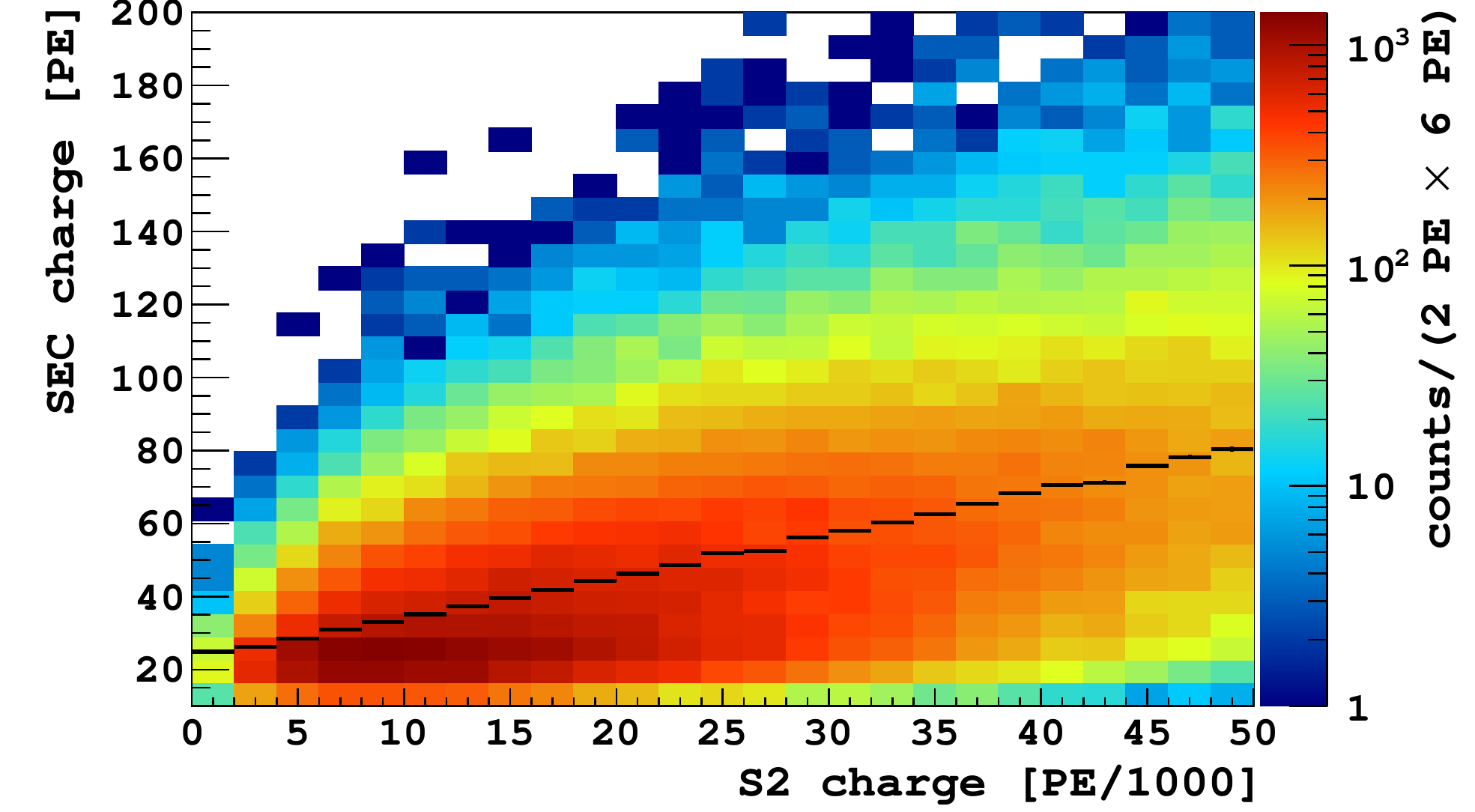}
 \caption{SEC charge vs. S2 charge distribution for S2-echo events. Overlaid is the profile histogram. A linear fit gives an intercept of $\sim$23.3 PE  and a slope of $\sim 1.2 \times 10^{-3}$.}
 \label{fig:Energy_correlation_S2_2}
\end{center}
\end{figure}

If our interpretation of the S2-echo events were correct, we should expect that, 
the larger the S2 charge, 
the larger the probability of inducing photoelectric emission from the cathode of more than one electron. 
Indeed this is observed   in \reffig{Energy_correlation_S2_2}, which shows the SEC charge vs. S2 charge distribution, for S2-echo events. Overlaid is the profile histogram, which clearly shows the expected correlation. 

\begin{figure}
\begin{center}
\includegraphics[width=\columnwidth]{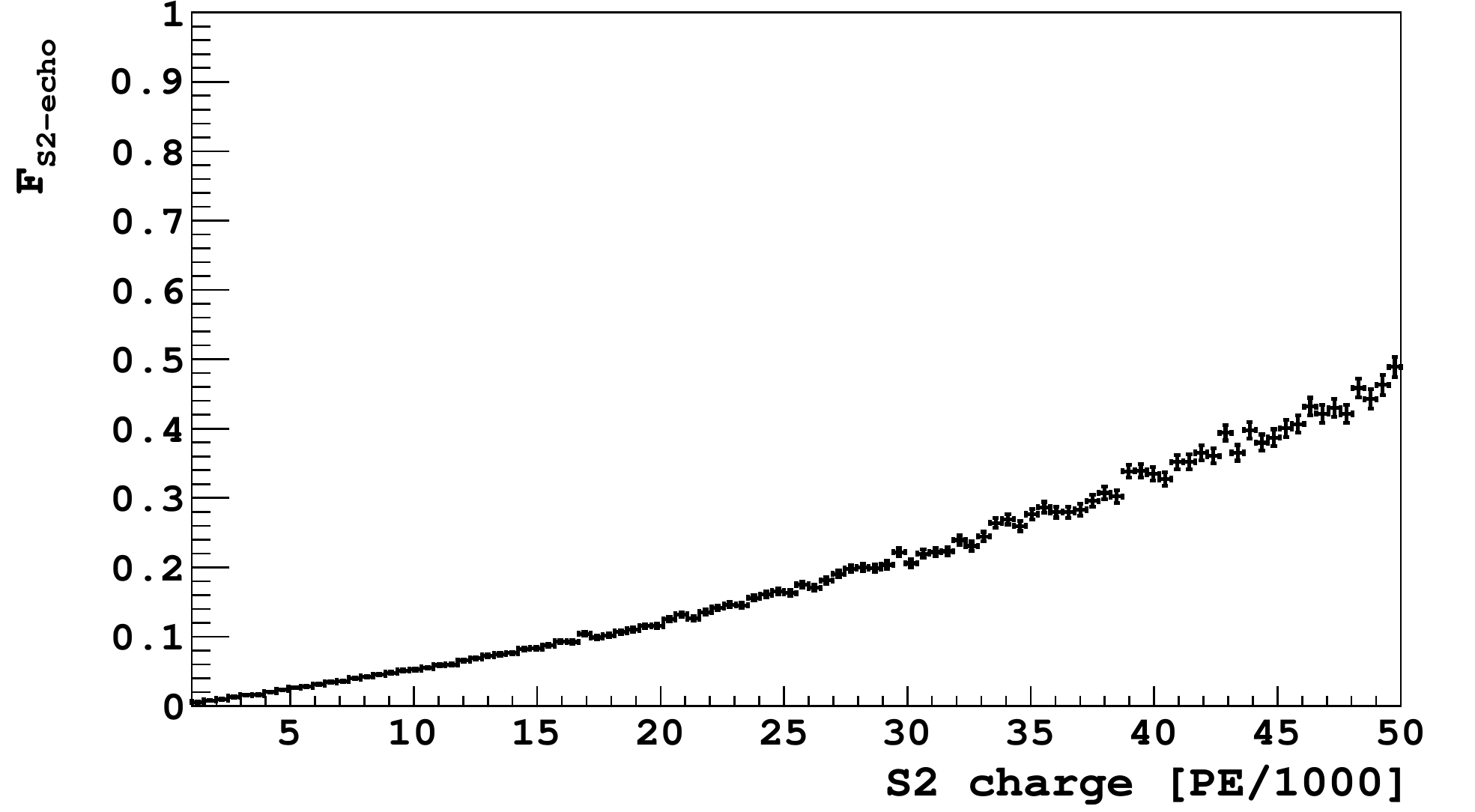}
\caption{Fraction of events containing S2-echoes vs. S2 charge, with $5\mus< \Delta t_{\mathrm{S2}-\mathrm{S1}}<45 \mus$.}
\label{fig:S2_ratio_distribution}
\end{center}
\end{figure}

We also expect that the probability of S2-echo events, regardless of  the SEC pulse  charge, increases with the S2 pulse charge.
Indeed, this is what is observed in  \reffig{S2_ratio_distribution}, which shows the fraction of events with an S2-echo
as a function of the S2 charge.
This fraction is found to monotonically increase with the S2 charge, leading to an event fraction of about 0.5 at the maximum S2 selected energy. 

 Therefore, S2-echo events, taking into account that we only select events  in the central top PMT out of 19, are very frequent. Indeed, they are present in almost every event, though in \DSfs\ data, due to the limited data acquisition  time window, most of the third pulses are not recorded on disk. 

\subsection{S1-echo events}
\label{sec:s1echo}
Another set of events  in~\reffig{SEC-S2} is clustered at  $\Delta t_{\mathrm{SEC}-\mathrm{S2}}$ between 50$~\mus$  and  $375~\mus$ and SEC charges peaking at $\sim$25 PE,  corresponding to the single-electron response. These events are well separated from those at SEC charges larger than a few 100 PEs, that are instead identified as S2 events from standard double-scatter $\gamma$-ray interactions in the detector. It can be noticed that, the pulse finder is not able to reconstruct SEC pulses below $\sim 100$ PE that are  less than  $\sim 40~\mu$s apart from an S2 pulse. 

\begin{figure}
\begin{center}
\includegraphics[width=\columnwidth]{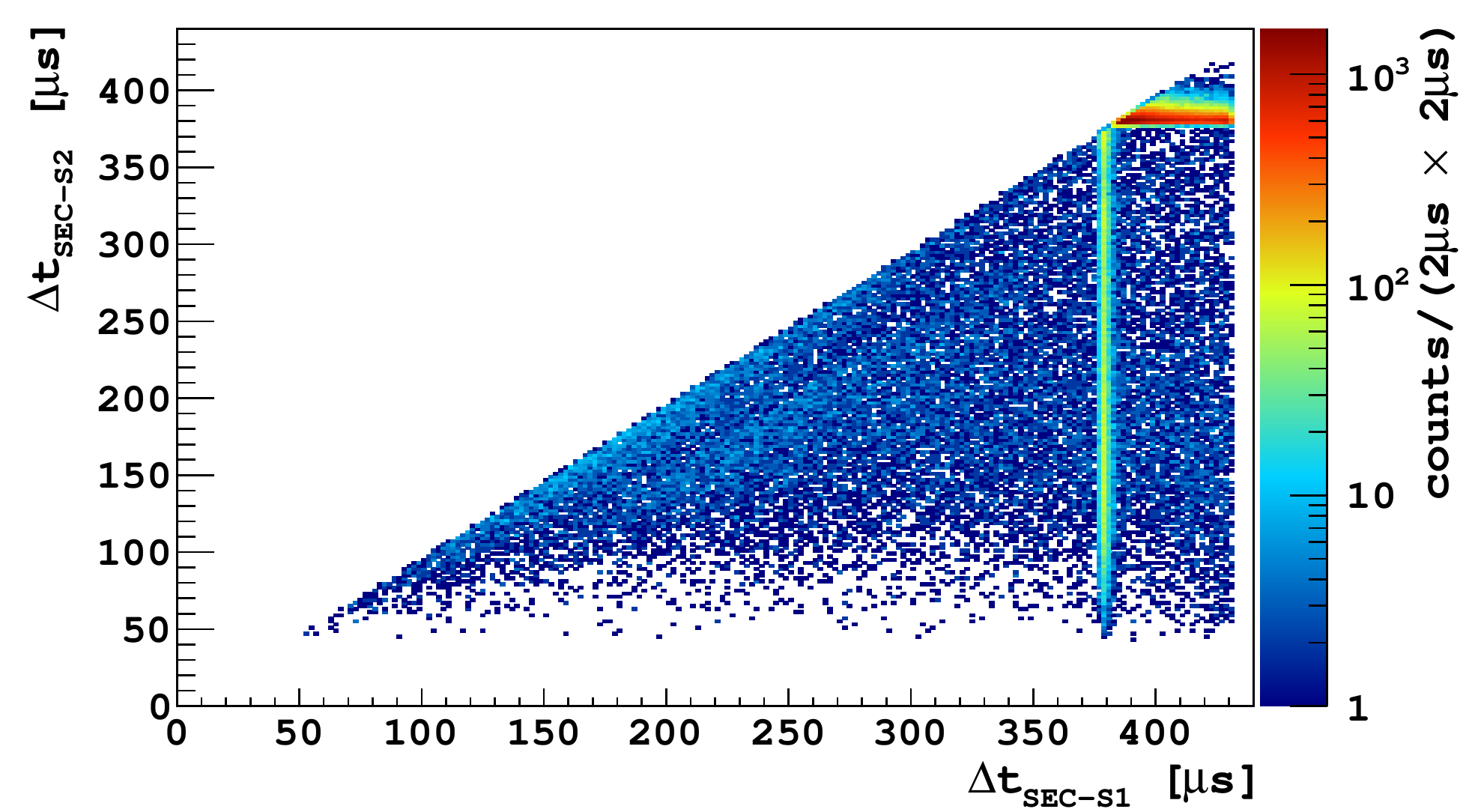}
  \caption{Time difference $\Delta t_{\mathrm{SEC}-\mathrm{S2}}$ vs.  time difference $\Delta t_{\mathrm{SEC}-\mathrm{S1}}$ distribution for events with  SEC$<$50 PE. 
  }
 \label{fig:SEC-S2_2}
\end{center}
\end{figure}

The origin of these  events can be understood from  \reffig{SEC-S2_2}, which  shows the distribution of $\Delta t_{\mathrm{SEC}-\mathrm{S2}}$ vs. $\Delta t_{\mathrm{SEC}-\mathrm{S1}}$, when selecting events with SEC$<$50 PE. 

\begin{figure}
\begin{center}
\includegraphics[width=\columnwidth]{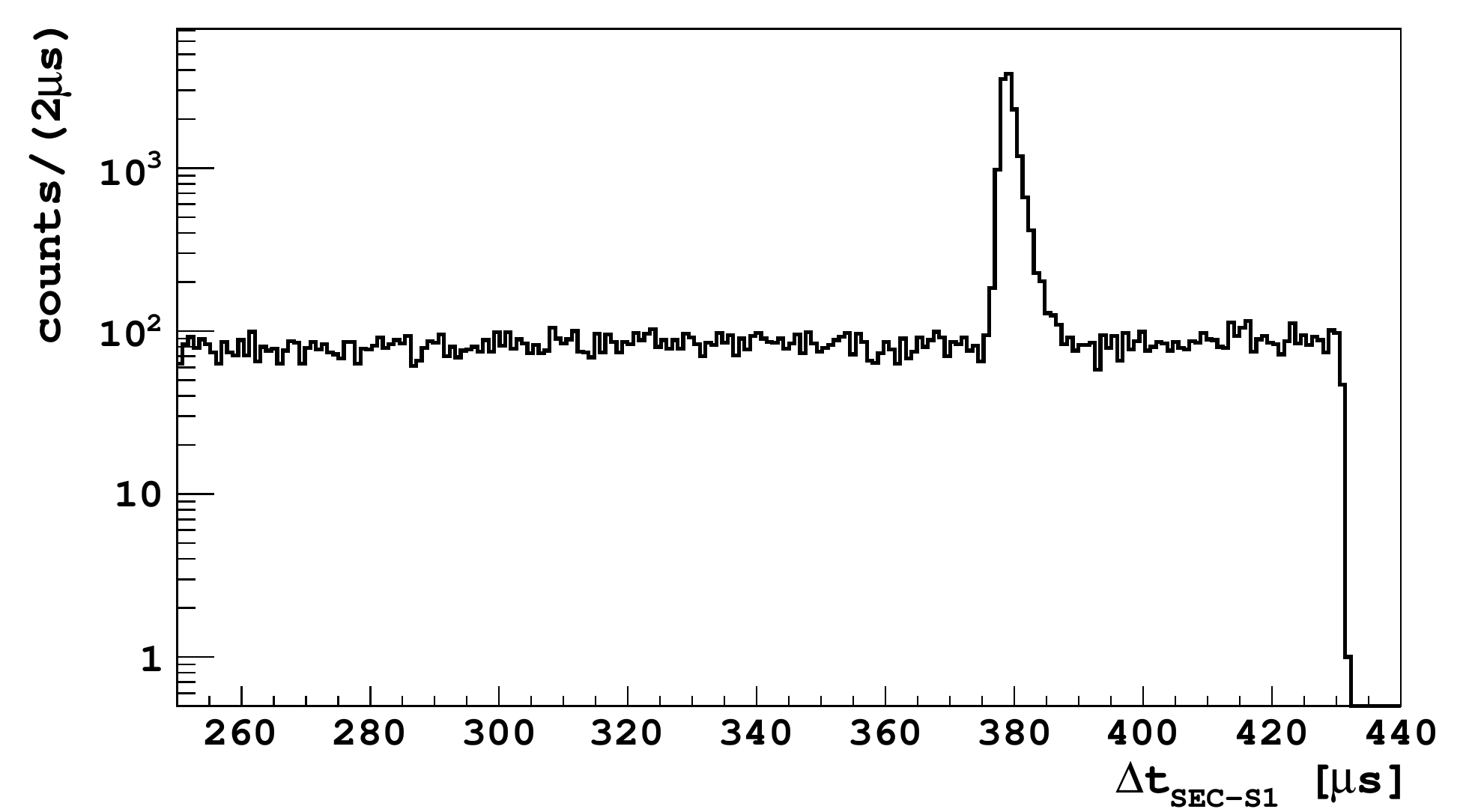}
  \caption{Time difference $\Delta t_{\mathrm{SEC}-\mathrm{S1}}$ distribution, for events with  SEC$<$50 PE and $\Delta t_{\mathrm{SEC}-\mathrm{S2}}<$ 350~$\mu$s. }
 \label{fig:monotime}
\end{center}
\end{figure}

Three event categories are clearly visible in the distribution: a horizontal band at $\Delta t_{\mathrm{SEC}-\mathrm{S2}} \sim 380 ~\mu\mathrm{s}$, corresponding to the S2-echo events discussed in \refsec{S2echo}, a continuum of events without   a specific time relation of the SEC with either S1 or S2, which will be discussed in \refsec{liquid}, and 
a vertical band, corresponding to $\Delta t_{\mathrm{SEC}-\mathrm{S1}} \sim 380 ~\mu\mathrm{s}$, about one maximum drift time after the S1 signal.  We interpret  these events, for $\Delta t_{\mathrm{SEC}-\mathrm{S2}} < 375 \mu\mathrm{s}$, as  photoelectric emissions from the cathode induced by S1 photons and call them {\em S1-echo} events. 
The  narrowness of the time distribution of the S1-echo events, shown in \reffig{monotime}, which displays $\Delta t_{\mathrm{SEC}-\mathrm{S1}}$ for events with  SEC$<$50 PE and $\Delta t_{\mathrm{SEC}-\mathrm{S2}}<$ 350 $\mu$s,  also confirms the expectation that, since  the electron extraction efficiency in the gas pocket is close to 100\%,  substantial delayed emission from the liquid surface on the scale of 10 to 100 $\mu$s  is excluded. 

\begin{figure}
\begin{center}
\includegraphics[width=\columnwidth]{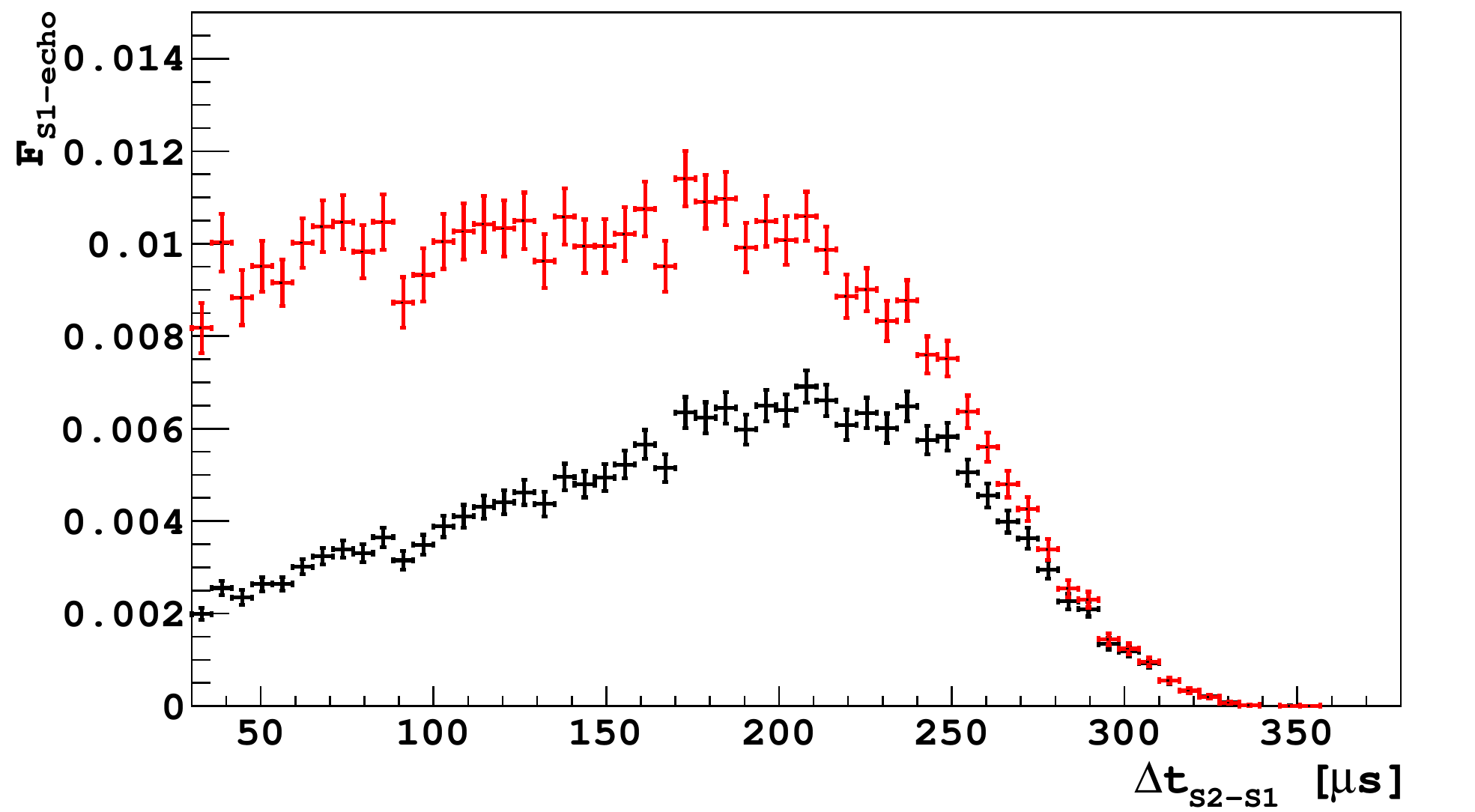}
 \caption{Fraction of events with an S1-echo, $F_{\mathrm{S1-echo}}$, vs. drift time $\Delta t_{\mathrm{S2}-\mathrm{S1}}$ (black dots) and after efficiency corrections (red dots), $F_{\mathrm{S1-echo}}^{\epsilon}$,  scaled by a factor 0.01.}
 \label{fig:S1_fraction}
\end{center}
\end{figure}

\reffiginitpar{S1_fraction} shows the fraction of events with an S1-echo, $F_{\mathrm{S1-echo}}$, vs. the drift time, $\Delta t_{\mathrm{S2}-\mathrm{S1}}$, defined as:
\begin{dmath}
F_{\mathrm{S1-echo}}(t_{\mathrm{drift}}) =  \frac{N_{\mathrm{S1-echo}}(t_{\mathrm{drift}})}  { {N}_{\mathrm{S2}}(t_{\mathrm{drift}})}, 
\label{eq:Fs1echodef} 
\end{dmath}
with $N_{\mathrm{S2}}(t_{\mathrm{drift}}) $ the selected total number of events (two pulse  and three pulse).
$F_{\mathrm{S1-echo}}$ rises up to about $250~\mus$, due to   acceptance effects,  whereas it      drops  at large $\Delta t_{\mathrm{S2}-\mathrm{S1}}$, when the time of the SEC become closer to the preceding S2. This drop   is due to a pulse finder inefficiency, likely the same  effect seen in \reffig{SEC-S2}, which tends to merge small signals with a preceding S2. Indeed, when for instance we select low energy events, such as  with S1$<$800 PE and S2$<$ 5000 PE,  we find that the drop at large $\Delta t_{\mathrm{S2}-\mathrm{S1}}$ only starts at  $\sim 300~\mus$. As a matter of fact, no tuning of the pulse finder algorithm was ever made to cope with this effect. 
We also tested the hypothesis that the drop could be  due to  the SEC being captured by the ion cloud of the S2 signal, by selecting events for which the S2 signal maximum is not in the central PMT. The corresponding distribution of \reffig{S1_fraction} does not change and, therefore, we discard this hypothesis.
The presence of a time gap between the S2 and the subsequent SEC is also visible in the continuum of  events at the bottom of \reffig{SEC-S2_2}.

\begin{figure}
\begin{center}
\includegraphics[width=\columnwidth]{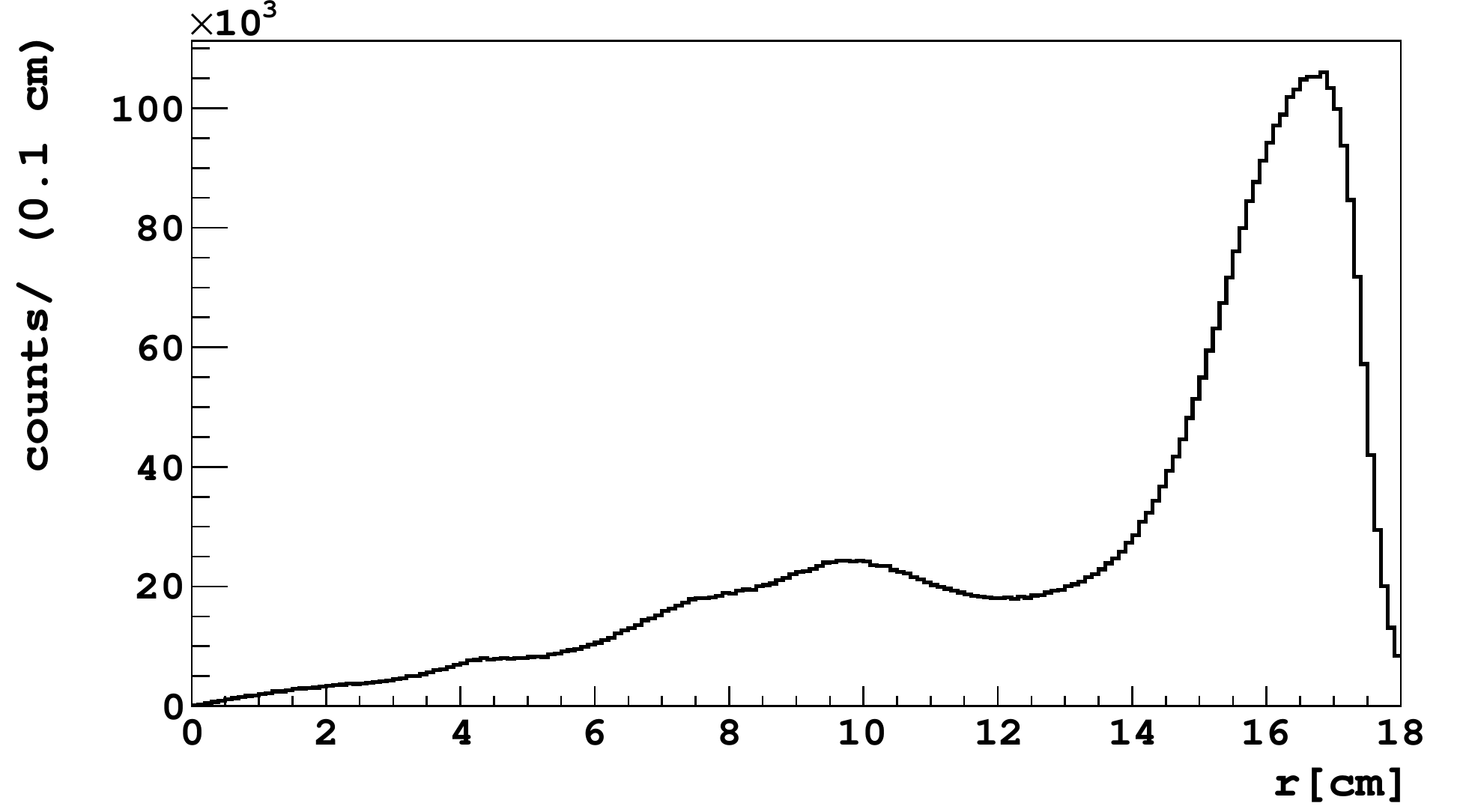}
 \caption{Radial distribution of S2 pulses for events  with  $\Delta t_{\mathrm{S2}-\mathrm{S1}}>$ 50 $\mu$s.}
 \label{fig:S2radial}
\end{center}
\end{figure}

The  efficiency for S1-echo events, ${\epsilon} (r,t_{\mathrm{drift}})$, was calculated with a toy Monte Carlo as  the fraction of  S1 UV photons, for which we assume 4$\pi$ emission at given $r$ and   $z$ position in the chamber corresponding to a given $t_{\mathrm{drift}}$, that hit a cathode area corresponding to the central PMT. In the following,  we made the simplifying assumption, true to a good approximation, that the event distribution in $t_{\mathrm{drift}}$ and $r$ factorize.  Then,  the average $\hat{\epsilon} (t_{\mathrm{drift}})$ is obtained by weighting  the efficiency ${\epsilon} (r,t_{\mathrm{drift}})$ by  the   radial distribution, $f(r)$, of the S2 pulses measured with data, as:

\begin{dmath}
  \hat{\epsilon} (t_{\mathrm{drift}})= \sum_{r} \epsilon (r,t_{\mathrm{drift}})   f(r),
\label{eq:3} 
\end{dmath}
The radial distribution of S2 events is shown in 
\reffig{S2radial} and is peaked at   large $r$ due to the material radioactivity.
For  $t_{\mathrm{drift}}<330~\mu s$, the calculated efficiency is a rising  function of $t_{\mathrm{drift}}$ and can be parametrized as:
\begin{dmath}
\hat{\epsilon} (t_{\mathrm{drift}})= 0.0072 \cdot e^{0.0024 \cdot t_{\mathrm{drift}}} - 0.0054.
\label{eq:effvsdrift} 
\end{dmath}
The fraction of events with an S1-echo vs. drift time,   after efficiency corrections, defined as
\begin{dmath}
F_{\mathrm{S1-echo}}^{\epsilon}(t_{\mathrm{drift}}) =  \frac{N_{\mathrm{S1-echo}}(t_{\mathrm{drift}})}    {\hat{\epsilon} (t_{\mathrm{drift}})  {N}_{\mathrm{S2}}(t_{\mathrm{drift}})},
\label{eq:Fs1echo} 
\end{dmath}
 is shown in red in \reffiginitpar{S1_fraction}. Below $\sim 200~\mu s$ we retrieve a flat distribution (the value of $F_{\mathrm{S1-echo}}^{\epsilon}$ may go above one since we consider  UV-photon emission in 4$\pi$). 

\begin{figure}
\begin{center}
\includegraphics[width=\columnwidth]{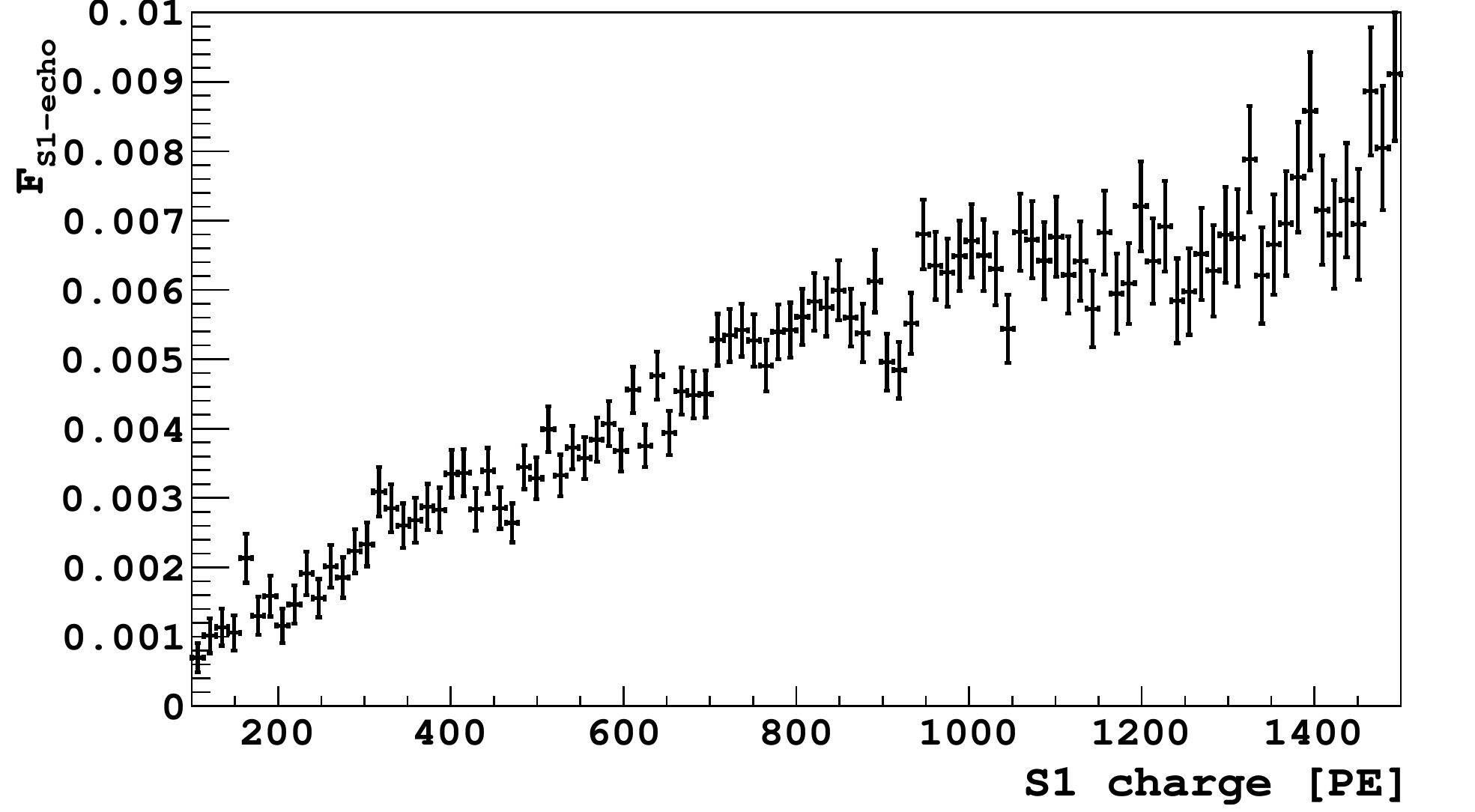}
\caption{Fraction of events with an S1-echoe vs. S1 charge, with $50~\mus< \Delta t_{\mathrm{S2}-\mathrm{S1}}<200~\mus$.}
\label{fig:S1_ratio_distribution}
\end{center}
\end{figure}

\begin{figure}
\begin{center}
\includegraphics[width=\columnwidth]{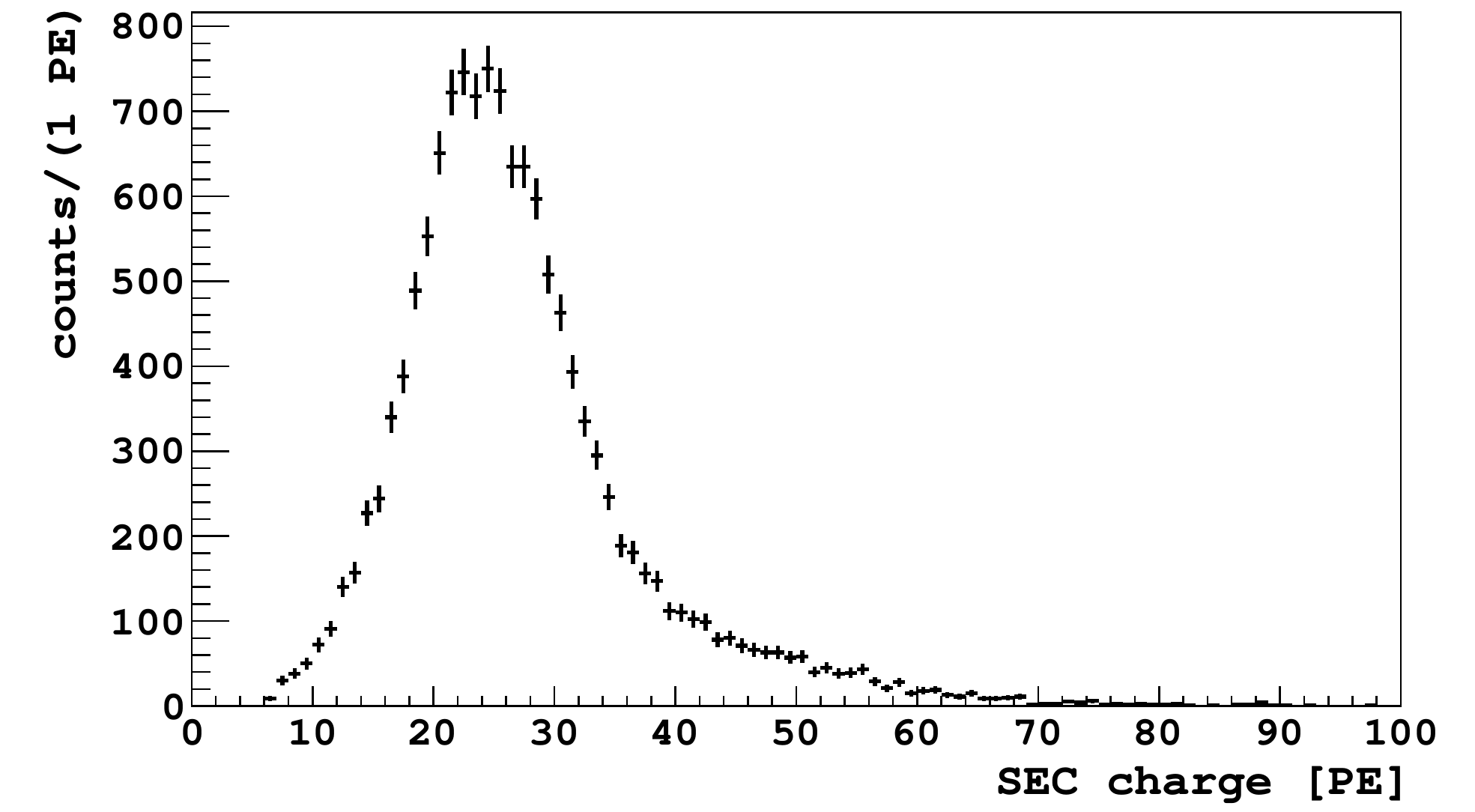}
 \caption{SEC charge spectrum for S1-echo events. }
 \label{fig:sec_charge_S1echo}
\end{center}
\end{figure}

By analogy with  \reffig{S2_ratio_distribution}, 
we show in \reffig{S1_ratio_distribution} the fraction of events containing S1-echoes vs. S1 charge, with $50~\mus< \Delta t_{\mathrm{S2}-\mathrm{S1}}<200~\mus$. Again larger S1 pulses have larger probability to produce also S1-echoes.
\reffiginitpar{sec_charge_S1echo} shows the SEC charge distribution for selected S1-echo events. The peak corresponding to one extracted electron can be clearly observed. A shoulder due to two extracted electrons can also be noticed.

\subsection{Calculation of the cathode quantum efficiency}

From the measured fraction of  both S1-echo and S2-echo events it is possible to estimate the quantum efficiency of the  cathode in liquid argon, $QE$, i.e the photoelectron emission probability per UV photon, $\gamma_{\mathsmaller{UV}}$, at the liquid argon emission wavelengths of $\sim$128~nm.


We select events with   $50~\mus< \Delta t_{\mathrm{S2}-\mathrm{S1}}<200~\mus$. Indeed,   we showed in \refsec{s1echo} that for these drift times we understand our efficiency corrections as we are able to retrieve a flat distribution as a function of $t_{\mathrm{drift}}$.  Then:
\begin{dmath}
  N_{\mathrm{S1-echo}}(t_{\mathrm{drift}}) =  \hat{\epsilon} (t_{\mathrm{drift}}) N_{\gamma_{\mathsmaller{UV}}} (t_{\mathrm{drift}})     
    QE_{S1},
\label{eq:ns1echo} 
\end{dmath}
where the number of UV photons is given by:
\begin{dmath}
   N_{\gamma_{\mathsmaller{UV}}} (t_{\mathrm{drift}}) =   N_{\mathrm{S2}}(t_{\mathrm{drift}})  \langle{S1}\rangle  / g_1.
\label{eq:ngamma} 
\end{dmath}
with  $g_1\sim 0.16$~PE/$\gamma_{\mathsmaller{UV}}$  \cite{Agnes:2017grb}  the collection efficiency of S1 photons generated in liquid argon, and  $\langle{S1}\rangle $ the S1 mean charge expressed in PE and assumed to be independent of $r$ and $t_{\mathrm{drift}}$. 


Since in the selected $\Delta t_{\mathrm{S2}-\mathrm{S1}}$ 
range $F^{\epsilon}_{\mathrm{S1-echo}}$ turns out to be  about constant (see \reffig{S1_fraction}), $\langle{F^{\epsilon}_{\mathrm{S1-echo}}}\rangle  \sim 1.0$, combining 
 \refeqn{Fs1echo}, \refeqn{ns1echo}, and \refeqn{ngamma}, we obtain
\begin{dmath}
         QE_{\mathrm{S1}} \sim \langle{F^{\epsilon}_{\mathrm{S1-echo}}}\rangle  \frac{g_1}{
    \langle{S1}\rangle 
    }
\label{eq:qe} 
\end{dmath}
Since    $\langle{S1}\rangle$  $\sim 730 $~PE, we obtain  $QE_{\mathrm{S1}} \sim 3 \times 10^{-4}/\gamma_{\mathsmaller{UV}}$.

  The S2-echo photons are induced by S2 signals. 
 Therefore,  only one value for  average geometric efficiency is needed, $\hat{\epsilon}_{S2}$, which corresponds to the value calculated from \refeqn{effvsdrift} at $t_{\mathrm{drift}}\sim 0$. The number of events with S2-echo is given by:
  \begin{dmath}
    N_{\mathrm{S2-echo}}(t_{\mathrm{drift}})  =  \hat{\epsilon}_{S2}    N_{\mathrm{S2}} (t_{\mathrm{drift}})
   \frac{ \langle{S2}\rangle  }{\langle{N_{el}}\rangle g_2}     QE_{S2},
\label{eq:1} 
\end{dmath}
where $\langle{N_{el}}\rangle$ is the average number of electrons per S2-echo event and 
 $g_2\sim 0.16$~PE/$\gamma_{\mathsmaller{UV}}$  \cite{Agnes:2017grb,Agnes:2021zyq}.
Since 
\begin{dmath}
\langle{N_{el}}\rangle=\langle{\mathrm{SEC}}\rangle/g
\end{dmath}
with $\langle{\mathrm{SEC}}\rangle$ the average SEC charge in PE and $g\sim 23\,\mathrm{PE}/e^-$  the photoelectric gain in the central PMT. Defining $K = {g}/{g_2}$ and taking the average
 of ${F_{\mathrm{S2-echo}}}$, defined in  \refeqn{defS2echo},  over the interval $5~\mus< \Delta t_{\mathrm{S2}-\mathrm{S1}}<45~\mus$, 
we obtain:
\begin{dmath}
    QE_{\mathrm{S2}} \sim \langle{F_{\mathrm{S2-echo}}}\rangle  
    \frac{1}{
    K    \hat \epsilon_{\mathrm{S2}}}  \frac{\langle{\mathrm{SEC}}\rangle }{\langle{S2}\rangle } 
\label{eq:qe} 
\end{dmath}
Now, from \reffig{S2/S1_echo_vs_z} we derive  $\langle{F_{\mathrm{S2-echo}}}\rangle \sim 0.055$, and we have 
  ${\langle{\mathrm{SEC}}\rangle }\sim$ 49~PE, ${\langle{S2}\rangle }\sim 23,430$~PE and  $ \hat\epsilon_{\mathrm{S2}}\sim 3\cdot 10^{-3}$.  Eventually, we obtain $QE_{\mathrm{S2}} \sim 4 \times 10^{-4}/\gamma_{\mathsmaller{UV}}$.

Both $QE_{\mathrm{S1}}$ and   $QE_{\mathrm{S2}}$ measurements, in fair agreement with each other,  are affected by  systematic uncertainties due  to the dependence  of both $g_1$ and $g_2$ on the interaction position in the detector, at most a 10-20\% effect, and  to the  geometric efficiency calculation.
Indeed, both $\hat{\epsilon} (t_{\mathrm{drift}})$ and  $\hat\epsilon_{\mathrm{S2}}$   were  calculated 
  by disregarding the number of photons hitting the cathode in the area between PMTs. An upper bound to the size of this effect 
   was evaluated by calculating the fraction of  the cathode surface covered by the PMTs divided by the number of the top array PMTs and it amounts to $\sim $10\%. 
    Rayleigh scattering was also not included in the  efficiency calculation.  An upper bound to the size of this effect could be obtained  by re-calculating $\hat{\epsilon} (t_{\mathrm{drift}})$ and  $\hat\epsilon_{\mathrm{S2}}$ with the inclusion in the toy Monte Carlo of  the Rayleigh scattering  probability
   for the UV-photons, with a  scattering length of $90$~cm \cite{Babicz_2020}, assuming  that every scattered photon is lost.
Eventually,  $\hat\epsilon_{\mathrm{S2}}$ would decrease   by $\sim $30\%, whereas  $\hat{\epsilon} (t_{\mathrm{drift}})$  by only  $\sim $15\%. 

 In the efficiency calculations we assumed no dependence on the angle of incidence on the cathode of the photoelectric efficiency, apart from the geometrical effects and  that UV light attenuation in liquid argon  is negligible. 

The  measured  absorption length of TPB at 128~nm is about 400~nm \cite{Benson:2017vbw}. Since this thickness is much smaller than the  few microns  of the TPB on the \DSfs\ cathode (see \refsec{detector}), most  photons are expected to give photoelectric effect in the TPB, and, therefore,  QE is an estimate of the  photoelectric quantum efficiency of the TPB, which is unmeasured so far.
It should also be noted that this may not be compared directly with a measurement in vacuum. Indeed, it is known that there could be a modification of the effective work function of the TPB by the electron affinity of the liquid argon, as  is expected for  liquid xenon \cite{PhysRevD.102.092004}.

\section{Bulk events}
\label{sec:liquid}

\subsection{Event features}
\label{sec:liquidf}
In addition to the S1-echo and S2-echo vertical and horizontal bands,    in \reffig{SEC-S2_2} there is also a continuum of events with no definite values of time differences of the SEC with either S1 or S2.
Since these events  follow in time the S2 signal,  we call them \textit{\stwoliquid} events.
%

\begin{figure}
\begin{center}
\includegraphics[width=\columnwidth]{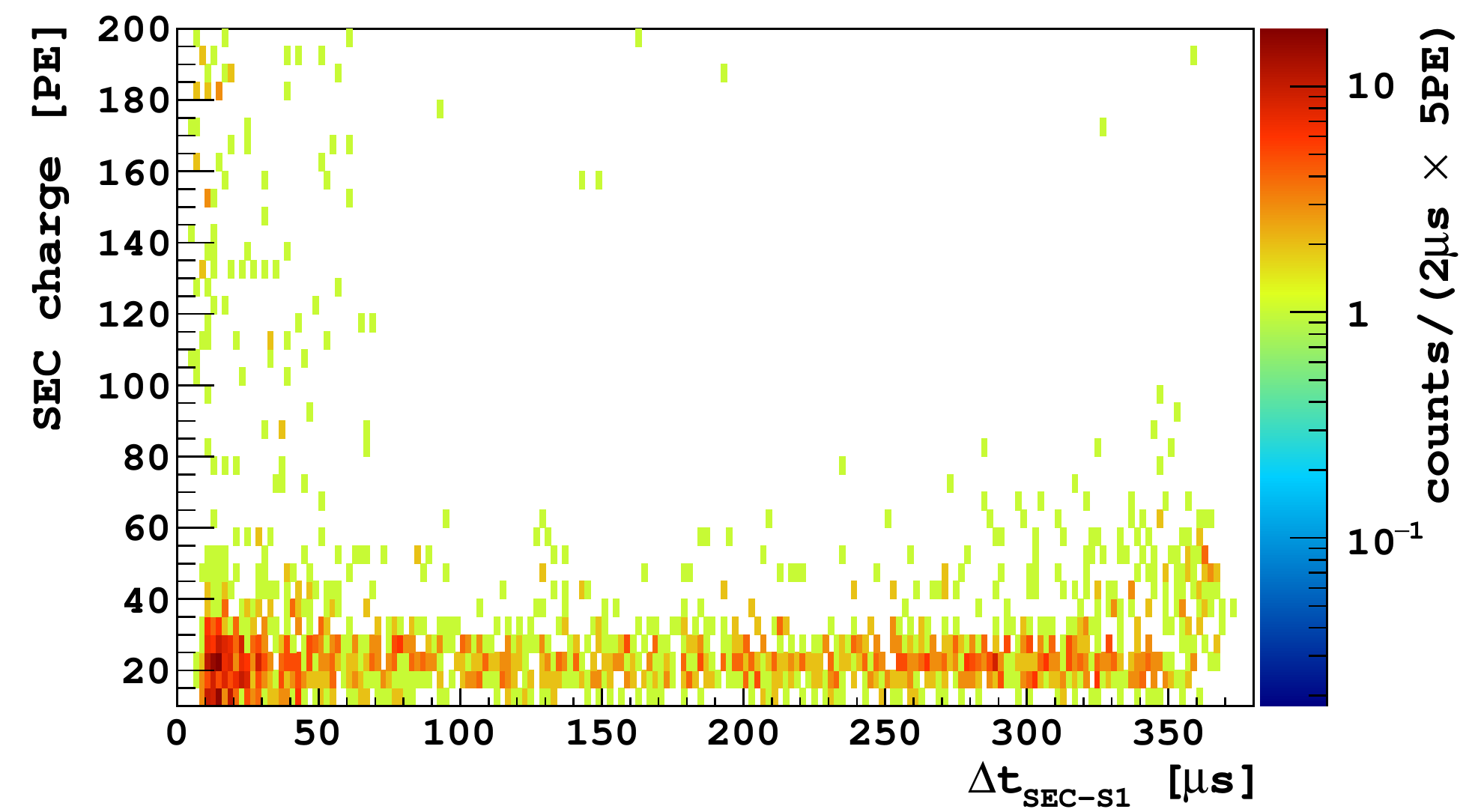}
 \caption{SEC charge vs. time difference $\Delta t_{\mathrm{SEC}-\mathrm{S1}}$ distribution, in events with the time sequence S1-SEC-S2.  }
 \label{fig:S1-SEC-S2}
\end{center}
\end{figure}

It is also possible to observe other events with no definite values of time difference of SEC with S1, by studying events with the time sequence S1-SEC-S2.
In \reffig{S1-SEC-S2}, we show  the SEC charge vs.  time difference $\Delta t_{\mathrm{SEC}-\mathrm{S1}}$ distribution.  
We notice that these events, apart from the first  30$~\mu$s,  are evenly distributed in time and with a SEC charge consistent with  being single electrons. We
call them \textit{\soneliquid} events. 

\subsection{Event selection summary}

The number of selected events after cuts,  N$_{ev}$, in the different categories studied in the previous sections is shown in \reftab{numbers}. We used different $\Delta t_{\mathrm{S2}-\mathrm{S1}}(\mu s)$ cuts for S1-echo compared to S2-echo,  to avoid  spill-over from S2-echo.
The cut applied to S2-echo and S2-bulk is applied to be  able to compare events where the  limited data-acquisition time window does not affect the selection of  S2-echo events.
S1-echo events are much more rare than S2-echo events, by  a factor of $\sim 70$ (taking also into account a factor $\sim 1.5$ in the  ratio of the geometric efficiencies, i.e.  the average value of $\hat{\epsilon} (t_{\mathrm{drift}})$, see \refeqn{effvsdrift}, and $\hat \epsilon_{\mathrm{S2}}$).

\begin{table}
\centering
\caption{Number of selected events after cuts,  N$_{ev}$, in the different categories and with given $\Delta t_{\mathrm{S2}-\mathrm{S1}}(\mu s)$ cut. }
\begin{tabular}{lcr}
    \hline

 {\textit{Category}} & $\Delta t_{\mathrm{S2}-\mathrm{S1}}(\mu s)$ & N$_{ev}$ \\ 
      
        \hline

       {S2-echo}& [5,45] &106964 \\
        {S1-echo}& [50,90] &2193  \\
 
        {\stwoliquid} & [5,45] & 8348\\
 
        {\soneliquid}& [50,90] &184  \\
        \hline
    \end{tabular}
 \label{tab:numbers}  
\end{table}

\subsection{Interpretation of bulk events}
\label{sec:S2interpretation}
\begin{figure}
\begin{center}
\includegraphics[width=\columnwidth]{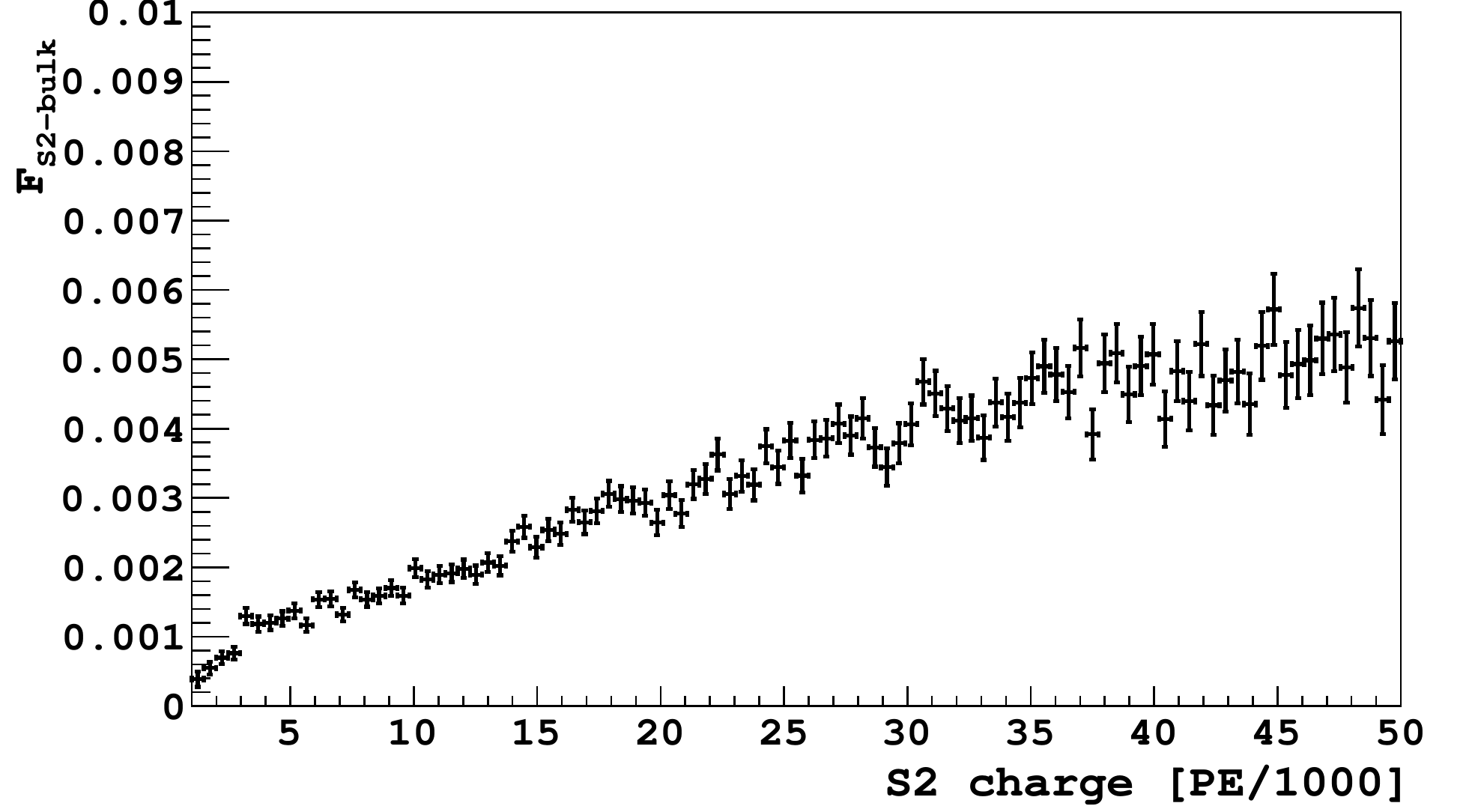}
\caption{Fraction of \stwoliquid\ events  vs. S2 charge, with $ \Delta t_{\mathrm{S2}-\mathrm{S1}}>50~\mus$.}
\label{fig:Sbulk_ratio_distribution}
\end{center}
\end{figure}

Understanding of the origin of \stwoliquid\ and \soneliquid\  events is not straightforward.  However, at least for the \stwoliquid\ events, we can say that their larger number compared to the \soneliquid\ ones suggests a link with S2 photons, given the much larger S2 pulse charge compared to S1.
However, while   events with $\Delta t_{\mathrm{SEC}-\mathrm{S1}}>380~\mu $s  can only be caused by S2, for events  with $\Delta t_{\mathrm{SEC}-\mathrm{S1}}<370~\mu$s 
we do not know  what fraction  is due to S1 or  S2. 
Another interesting observation for \stwoliquid\ events is that also the fraction of \stwoliquid\ events increases with S2 charge, as shown in \reffig{Sbulk_ratio_distribution}.
%

A candidate explanation for the \stwoliquid\ events is the photoionization by S2 photons (or S1 photons) of   contaminants. One example is a contaminant  which has previously captured an electron   during  a past event, such as for instance $\textrm{O}_2^-$, which  has a relatively low  ionization energy, namely below  the 9.6 eV  of liquid  argon emission. The photoionization of neutral  molecules such as $\textrm{O}_2$ or $\textrm{H}_2\textrm{O}$ is less likely since the first ionization energy is larger than 9.6 eV.  Another possible explanation for the \stwoliquid\ events is photoionization of TPB dissolved in the liquid.
%

To test the  contaminant hypothesis,  we analyzed a set of data taken in a time period of five days in July 2015, when the getter was turned off for maintenance. As a matter of fact, during this period, we expect an increase of contaminants and, hence, an increase of photoionization in the bulk. 
Indeed, during the same time period, as described in a previous \DSfs\ paper \cite{Agnes:2018fg}, we observed  a five-fold increase of isolated, i.e. far in time from a standard  event,  single electrons.
However, with the data of this paper, we measure  that the number  of \stwoliquid\ events increased only by $\sim$35\%, with respect to the number of S2-echo events,  indicating a somewhat different mechanism for the production of S2-bulk events from that of single isolated electrons.
We disfavor the possible interpretation of the \stwoliquid\ events as being due to recombination or molecule de-excitation since these mechanisms are not expected to yield  electrons.

From the number of \stwoliquid\ events,  we  derive the probability of photoelectric extraction from the liquid per unit length, $PEP_{S2}$, for $5~\mu s< \Delta t_{\mathrm{S2}-\mathrm{S1}}<45~\mu s$, as 
 \begin{dmath}
    N_{\mathrm{S2-bulk}}  =  \sum_r {L(r)   N_{\mathrm{S2}}(r)}  
    \frac{\langle{S2}\rangle }{ g_2}    PEP_{S2},
\label{eq:14} 
\end{dmath}
where ${L(r)}$ is the path length,  inside a cylinder of diameter equal to that of the central PMT and height equal to the maximum TPC drift length, of an S2 photon generated at the radial distance $r$, $N_{\mathrm{S2}}(r)$ the  selected total number of events (two pulse and three pulse) vs $r$, and $\langle{S2}\rangle \sim 23,430$~PE.

Eventually, the measured $PEP_{S2}$ is  $\sim 3 \times 10^{-6}$ $e^-/\gamma_{\mathsmaller{UV}}/\mathrm{m}$.

Due to the SEC pulse selection requirement of having the signal maximum  in the central PMT, we tend to rule out the interpretation of \stwoliquid\ events  as photoelectric emissions from the lateral walls. 

To fully understand the origin of these events, further experimental investigation is needed.

\begin{figure}
\begin{center}
\includegraphics[width=\columnwidth]{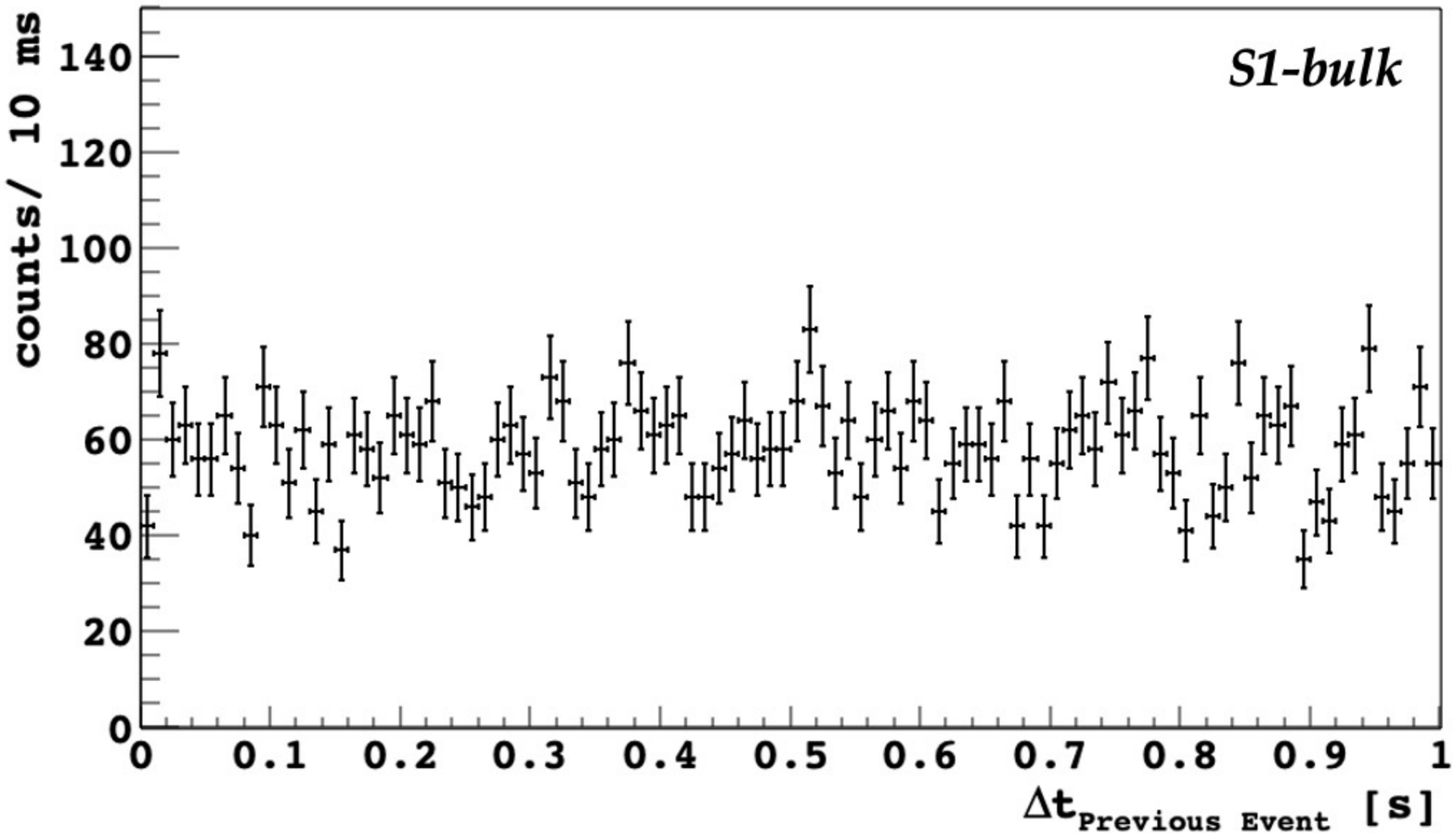}
 \caption{Distribution of the time difference of \soneliquid\ events with  any previous event in a time window of 10 s, for SEC $<$50 PE. No specific selection on the previous events is applied.  }
 \label{fig:S1before}
\end{center}
\end{figure}

As far as the \soneliquid\ events are concerned, there are two possible interpretations for the origin of the SECs: either they  are due to S1 or they  are   remnants from  previous events, e.g. electrons captured by some electronegative impurity and then released accidentally in the time window between  S1 and  S2.
To test this hypothesis, we looked at a possible time correlation with the previous events. The  time difference of \soneliquid\ events with  any previous event in a time window of 10~s (out of which we only display 1~s), for SEC$<$50~PE, is shown in \reffig{S1before}. 
No specific selection to the previous events is applied and the time of the events is defined here as the trigger time. According to a toy Monte Carlo simulation that we have performed, a time  correlated component would show up as an exponential rise towards zero time. Therefore, from the  experimental distribution, we can exclude, at present level of statistics, a  correlated component with $\tau\gtrapprox 20$~ms.

A larger than average number of events per unit $\Delta t_{\mathrm{SEC}-\mathrm{S1}}$, and  also with a higher SEC charge, is observed in \reffig{S1-SEC-S2} at small values of $\Delta t_{\mathrm{SEC}-\mathrm{S1}}$, below 30$~\mu$s. A possible interpretation of these events is the  photoionization of the extraction  grid by S1 signals, as observed also with the   LUX detector \cite{PhysRevD.102.092004}.

The ratio of number of   \soneliquid\ to  \stwoliquid\ events also follows roughly  the ratio between the S2 and S1 pulse charges, bearing in mind that  we do not know what fraction of \stwoliquid\ events is related to S1 and the inefficiency due to somewhat different selection cuts. 
\section{Discussion}
\label{sec:xenon}
We observed several categories of single isolated electrons in association with standard  scintillation-ionization S1-S2 signals with the \DSfs\  LAr TPC.
Since this is the first study in an argon detector, it is interesting to compare our results to the abundant literature available with xenon detectors.
The LUX Collaboration \cite{PhysRevD.102.092004} reports about  four kinds of  phenomena, three of which   are also observed in \DSfs: {\em photoionization electrons that are detected within hundreds of microseconds after the S1 and S2 pulses}, which are  described in this  paper, {\em delayed emission of individual electrons at the millisecond-to-second scale}, and  {\em electron emission that appears independent of prior interactions}, that are briefly discussed in Ref.~\cite{Agnes:2018fg} and will be treated in more detail in a upcoming \DSfs\ publication. On the contrary,  we do not observe in \DSfs\  {\em clustered electron emission that occurs within tens of milliseconds after S2}. 

As far as photoionization electrons are concerned, their occurrence in xenon detectors has both similarities and differences with our findings. 

We expect differences at least for the following reasons. The liquid to gas extraction efficiency with the  electric fields used by the experiments is about 100\% for argon, while it is only $\sim $50\% for xenon \cite{Gushchin:1982b}, leading to potential electron trapping at the surface and, therefore, delayed electron emission. The measured electron lifetime in \DSfs\ is much larger than both the lifetime measured by LUX (by more than a factor of 10) and  the \DSfs\ maximum drift time (by a factor of 30). Therefore,  electron capture by impurities  during the drift is expected to be much less of a  relevant issue in \DSfs. This also implies  that for \DSfs\ no efficiency correction is needed for S2 vs drift time.
Another relevant difference is that the \DSfs\ cathode and anode planes are continuous planes, with the surface facing the active volume coated with ITO and TPB, whereas LUX uses metal grids and  no wavelength shifting of the light.

S1-echo and S2-echo events are observed both  with xenon detectors,  e.g. LUX and XENON100 \cite{Aprile:2013blg}  and with  \DSfs. The quantum efficiency of TPB in \DSfs\ and of metal grids in LUX were measured. In both experiments they were calculated with both   S1 and S2 photons and the results agreed in both cases within a  factor of two.  

In both LUX and \DSfs\ we observe photoionization events from the extraction grid, right below the gas-liquid interface.

Both \soneliquid\ and \stwoliquid\ events   are observed by both LUX and \DSfs\ Collaborations. 
Interesting considerations about the origin of \stwoliquid\ events in LUX were obtained from the $\Delta t_{\mathrm{S2}-\mathrm{S1}}$ distribution.  Unfortunately, due to the limited time window of our data acquisition, we did not record on disk all \stwoliquid\ events. Indeed, as shown  in \reffig{SEC-S2_2}, all events above $\Delta t_{\mathrm{SEC}-\mathrm{S1}}=430~\mus$ are cut out and, therefore, a $\Delta t_{\mathrm{S2}-\mathrm{S1}}$ distribution (constant values of $\Delta t_{\mathrm{S2}-\mathrm{S1}}$ are diagonal lines with unit slope  in the figure) would be biased by the time acceptance cut.  Nonetheless,  it can be noticed that the \stwoliquid\ event density is quite constant (even  in linear scale) and, therefore, no significant dependence of the number of events on the interaction depth is observed.

\soneliquid\ and \stwoliquid\ events are attributed by LUX  to the photoionization of impurities, more likely neutral molecules than negative ions, dissolved in liquid xenon.
The hypothesis of photoionization in the liquid xenon was also suggested by the  XENON-100 \cite{Aprile:2013blg} Collaboration, which showed a correlation of the rate with the electron lifetime, and ZEPLIN-II Collaborations \cite{Edwards:2007nj}.
We also observed a correlation with the impurity concentration since we observed of a  rate increase during a  period of time  with the getter switched off. However, 
our understanding of the origin of these events is only partial and inconclusive. 

Both LUX and \DSfs\  measured the   probability of photoelectric emission per unit length in the bulk. LUX measured (5-20)$\times 10^{-5} e^-/\gamma_{\mathsmaller{UV}}/\mathrm{m}$  while \DSfs\ with \stwoliquid\ events measured $\sim$ 3$\times 10^{-6} e^-/\gamma_{\mathsmaller{UV}}/\mathrm{m}$. The smaller value found with \DSfs\ by more than a factor of 10, together with the much larger electron lifetime and  the larger photon energy in \DSfs\, coherently hints toward  a lower concentration of contaminants in \DSfs.
In both experiments, the identification of the impurity molecule was not possible and is left for future experimental work.

The S1-echo and  S2-echo events  observed with \DSfs\ are most probably going to be present also in \DSks, given  that the same   wavelength shifter is going to  be  deposited on the cathode. Since the aspect ratio of the two TPCs is about the same, the number of expected echo events is going to scale with   the ratio of  background rates. 
For the  \stwoliquid\ events, there is going to be a factor of 5 more on top of the background rate factor, due to the larger drift length of \DSks,  were the level of contaminants is assumed to be the same.

\section*{Acknowledgements}
The DarkSide Collaboration offers its profound gratitude to the LNGS and its staff for their invaluable technical and logistical support. We also thank the Fermilab Particle Physics, Scientific, and Core Computing Divisions. Construction and operation of the DarkSide-50 detector was supported by the U.S. National Science Foundation (NSF) (Grants No. PHY-0919363, No. PHY-1004072, No. PHY-1004054, No. PHY-1242585, No. PHY-1314483, No. PHY-1314501, No. PHY-1314507, No. PHY-1352795, No. PHY-1622415, and associated collaborative grants No. PHY-1211308 and No. PHY-1455351), the Italian Istituto Nazionale di Fisica Nucleare, the U.S. Department of Energy (Contracts No. DE-FG02-91ER40671, No. DEAC02-07CH11359, and No. DE-AC05-76RL01830), the Polish NCN (Grant No. UMO-2014/15/B/ST2/02561) and the Foundation for Polish Science (Grant No. Team2016-2/17). We also acknowledge financial support from the French Institut National de Physique Nucl\'eaire et de Physique des Particules (IN2P3) and the UnivEarthS LabEx program (Grants No. ANR-10-LABX-0023 and No. ANR-18-IDEX-0001),  from the São Paulo Research Foundation (FAPESP) (Grant No. 2016/09084-0), from the Interdisciplinary Scientific and Educational School of Moscow University ``Fundamental and Applied Space Research'', and from IRAP AstroCeNT funded by FNP from ERDF. Isotopes used in this research were supplied by the United States Department of Energy Office of Science by the Isotope Program in the Office of Nuclear Physics.


\bibliographystyle{ds}
\bibliography{ds,wal}

\begin{thebibliography}{23}
\expandafter\ifx\csname natexlab\endcsname\relax\def\natexlab#1{#1}\fi
\expandafter\ifx\csname bibnamefont\endcsname\relax
  \def\bibnamefont#1{#1}\fi
\expandafter\ifx\csname bibfnamefont\endcsname\relax
  \def\bibfnamefont#1{#1}\fi
\expandafter\ifx\csname citenamefont\endcsname\relax
  \def\citenamefont#1{#1}\fi
\expandafter\ifx\csname url\endcsname\relax
  \def\url#1{\texttt{#1}}\fi
\expandafter\ifx\csname urlprefix\endcsname\relax\def\urlprefix{URL }\fi
\providecommand{\bibinfo}[2]{#2}
\providecommand{\eprint}[2][]{\url{#2}}

\bibitem[{\citenamefont{Billard et~al.}(2014)\citenamefont{Billard,
  Figueroa-Feliciano, and Strigari}}]{NF}
\bibinfo{author}{\bibfnamefont{J.}~\bibnamefont{Billard}},
  \bibinfo{author}{\bibfnamefont{E.}~\bibnamefont{Figueroa-Feliciano}},
  \bibnamefont{and} \bibinfo{author}{\bibfnamefont{L.}~\bibnamefont{Strigari}},
   \href{http://dx.doi.org/10.1103/physrevd.89.023524}{\bibinfo{journal}{Phys.
  Rev. D} \textbf{\bibinfo{volume}{89}}, \bibinfo{pages}{023524}\bibinfo{year}{
  (\bibinfo{year}{2014})}}.

\bibitem[{\citenamefont{Agnes et~al.}(2018{\natexlab{a}})}]{Agnes:2018fg}
\bibinfo{author}{\bibfnamefont{P.}~\bibnamefont{Agnes}}  \bibnamefont{et~al.}
  (\bibinfo{affiliation}{The DarkSide Collaboration}),
  \href{http://dx.doi.org/10.1103/PhysRevLett.121.081307}{\bibinfo{journal}{Phys.
  Rev. Lett.} \textbf{\bibinfo{volume}{121}},
  \bibinfo{pages}{081307}\bibinfo{year}{
  (\bibinfo{year}{2018}{\natexlab{a}})}}.

\bibitem[{\citenamefont{Agnes et~al.}(2018{\natexlab{b}})}]{Agnes:2018ft}
\bibinfo{author}{\bibfnamefont{P.}~\bibnamefont{Agnes}}  \bibnamefont{et~al.}
  (\bibinfo{affiliation}{The DarkSide Collaboration}),
  \href{http://dx.doi.org/10.1103/PhysRevLett.121.111303}{\bibinfo{journal}{Phys.
  Rev. Lett.} \textbf{\bibinfo{volume}{121}},
  \bibinfo{pages}{111303}\bibinfo{year}{
  (\bibinfo{year}{2018}{\natexlab{b}})}}.

\bibitem[{\citenamefont{Aalseth et~al.}(2018)}]{Aalseth:2018gq}
\bibinfo{author}{\bibfnamefont{C.~E.} \bibnamefont{Aalseth}}
  \bibnamefont{et~al.} (\bibinfo{affiliation}{The DarkSide Collaboration}),
  \href{http://dx.doi.org/10.1140/epjp/i2018-11973-4}{\bibinfo{journal}{Eur.
  Phys. J. Plus} \textbf{\bibinfo{volume}{133}},
  \bibinfo{pages}{131}\bibinfo{year}{ (\bibinfo{year}{2018})}}.

\bibitem[{\citenamefont{Agnes et~al.}(2018{\natexlab{c}})}]{Agnes:2018hvf}
\bibinfo{author}{\bibfnamefont{P.}~\bibnamefont{Agnes}} \bibnamefont{et~al.},
  \href{http://dx.doi.org/10.1016/j.nima.2018.06.077}{\bibinfo{journal}{Nucl.
  Instrum. Meth. A} \textbf{\bibinfo{volume}{904}},
  \bibinfo{pages}{23}\bibinfo{year}{ (\bibinfo{year}{2018}{\natexlab{c}})}}.

\bibitem[{\citenamefont{Boulay and Hime}(2006)}]{Boulay:2006hu}
\bibinfo{author}{\bibfnamefont{M.~G.} \bibnamefont{Boulay}} \bibnamefont{and}
  \bibinfo{author}{\bibfnamefont{A.}~\bibnamefont{Hime}},
  \href{http://dx.doi.org/10.1016/j.astropartphys.2005.12.009}{\bibinfo{journal}{Astropart.
  Phys.} \textbf{\bibinfo{volume}{25}}, \bibinfo{pages}{179}\bibinfo{year}{
  (\bibinfo{year}{2006})}}.

\bibitem[{\citenamefont{Hitachi et~al.}(1983)}]{PhysRevB.27.5279}
\bibinfo{author}{\bibfnamefont{A.}~\bibnamefont{Hitachi}}
  \bibnamefont{et~al.},
  \href{http://dx.doi.org/10.1103/PhysRevB.27.5279}{\bibinfo{journal}{Phys.
  Rev. B} \textbf{\bibinfo{volume}{27}}, \bibinfo{pages}{5279}\bibinfo{year}{
  (\bibinfo{year}{1983})}}.

\bibitem[{\citenamefont{Akerib et~al.}(2020)}]{PhysRevD.102.092004}
\bibinfo{author}{\bibfnamefont{D.~S.} \bibnamefont{Akerib}}
  \bibnamefont{et~al.},
  \href{http://dx.doi.org/10.1103/PhysRevD.102.092004}{\bibinfo{journal}{Phys.
  Rev. D} \textbf{\bibinfo{volume}{102}},
  \bibinfo{pages}{092004}\bibinfo{year}{ (\bibinfo{year}{2020})}}.

\bibitem[{\citenamefont{Edwards et~al.}(2008)}]{Edwards:2007nj}
\bibinfo{author}{\bibfnamefont{B.}~\bibnamefont{Edwards}} \bibnamefont{et~al.},
   \href{http://dx.doi.org/10.1016/j.astropartphys.2008.06.006}{\bibinfo{journal}{Astropart.
  Phys.} \textbf{\bibinfo{volume}{30}}, \bibinfo{pages}{54}\bibinfo{year}{
  (\bibinfo{year}{2008})}}.

\bibitem[{\citenamefont{Aprile et~al.}(2014)}]{Aprile:2013blg}
\bibinfo{author}{\bibfnamefont{E.}~\bibnamefont{Aprile}} \bibnamefont{et~al.},
  \href{http://dx.doi.org/10.1088/0954-3899/41/3/035201}{\bibinfo{journal}{J.
  Phys. G} \textbf{\bibinfo{volume}{41}},
  \bibinfo{pages}{035201}\bibinfo{year}{ (\bibinfo{year}{2014})}}.

\bibitem[{\citenamefont{Akimov et~al.}(2016)}]{Akimov_2016}
\bibinfo{author}{\bibfnamefont{D.}~\bibnamefont{Akimov}}  \bibnamefont{et~al.},
   \href{http://dx.doi.org/10.1088/1748-0221/11/03/c03007}{\bibinfo{journal}{Journal
  of Instrumentation} \textbf{\bibinfo{volume}{11}},
  \bibinfo{pages}{C03007}\bibinfo{year}{ (\bibinfo{year}{2016})}}.

\bibitem[{\citenamefont{Agnes et~al.}(2018{\natexlab{d}})}]{Agnes:2018ep}
\bibinfo{author}{\bibfnamefont{P.}~\bibnamefont{Agnes}}  \bibnamefont{et~al.}
  (\bibinfo{affiliation}{The DarkSide Collaboration}),
  \href{http://dx.doi.org/10.1103/PhysRevD.98.102006}{\bibinfo{journal}{Phys.
  Rev. D} \textbf{\bibinfo{volume}{98}}, \bibinfo{pages}{102006}\bibinfo{year}{
  (\bibinfo{year}{2018}{\natexlab{d}})}}.

\bibitem[{\citenamefont{Acosta-Kane et~al.}(2008)}]{AcostaKane:2008im}
\bibinfo{author}{\bibfnamefont{D.}~\bibnamefont{Acosta-Kane}}
  \bibnamefont{et~al.},
  \href{http://dx.doi.org/10.1016/j.nima.2007.12.032}{\bibinfo{journal}{Nucl.
  Inst. Meth. A} \textbf{\bibinfo{volume}{587}},
  \bibinfo{pages}{46}\bibinfo{year}{ (\bibinfo{year}{2008})}}.

\bibitem[{\citenamefont{Xu et~al.}(2015)}]{Xu:2015do}
\bibinfo{author}{\bibfnamefont{J.}~\bibnamefont{Xu}}  \bibnamefont{et~al.},
  \href{http://dx.doi.org/10.1016/j.astropartphys.2015.01.002}{\bibinfo{journal}{Astropart.
  Phys.} \textbf{\bibinfo{volume}{66}}, \bibinfo{pages}{53}\bibinfo{year}{
  (\bibinfo{year}{2015})}}.

\bibitem[{\citenamefont{Bondar et~al.}(2009)}]{Bondar_2009}
\bibinfo{author}{\bibfnamefont{A.}~\bibnamefont{Bondar}}  \bibnamefont{et~al.},
   \href{http://dx.doi.org/10.1088/1748-0221/4/09/p09013}{\bibinfo{journal}{Journal
  of Instrumentation} \textbf{\bibinfo{volume}{4}},
  \bibinfo{pages}{P09013}\bibinfo{year}{ (\bibinfo{year}{2009})}}.

\bibitem[{\citenamefont{Gushchin et~al.}(1982)\citenamefont{Gushchin, Kruglov,
  and Obodovskil}}]{Gushchin:1982b}
\bibinfo{author}{\bibfnamefont{E.~M.} \bibnamefont{Gushchin}},
  \bibinfo{author}{\bibfnamefont{A.~A.} \bibnamefont{Kruglov}},
  \bibnamefont{and} \bibinfo{author}{\bibfnamefont{I.~M.}
  \bibnamefont{Obodovskil}},
  \href{http://dx.doi.org/10.1088/1748-0221/16/07/p07014}{\bibinfo{journal}{Zh.Eksp.Teor.Fi.}
  \textbf{\bibinfo{volume}{82}}, \bibinfo{pages}{1485}\bibinfo{year}{
  (\bibinfo{year}{1982})}}.

\bibitem[{\citenamefont{Agnes et~al.}(2017{\natexlab{a}})}]{Agnes_2017}
\bibinfo{author}{\bibfnamefont{P.}~\bibnamefont{Agnes}}  \bibnamefont{et~al.},
  \href{http://dx.doi.org/10.1088/1748-0221/12/12/p12011}{\bibinfo{journal}{Journal
  of Instrumentation} \textbf{\bibinfo{volume}{12}},
  \bibinfo{pages}{P12011}\bibinfo{year}{
  (\bibinfo{year}{2017}{\natexlab{a}})}}.

\bibitem[{\citenamefont{Agnes et~al.}(2015)}]{Agnes:2015gu}
\bibinfo{author}{\bibfnamefont{P.}~\bibnamefont{Agnes}}  \bibnamefont{et~al.}
  (\bibinfo{affiliation}{The DarkSide Collaboration}),
  \href{http://dx.doi.org/10.1016/j.physletb.2015.03.012}{\bibinfo{journal}{Phys.
  Lett. B} \textbf{\bibinfo{volume}{743}}, \bibinfo{pages}{456}\bibinfo{year}{
  (\bibinfo{year}{2015})}}.

\bibitem[{\citenamefont{Acciarri et~al.}(2010)}]{Acciarri_2010}
\bibinfo{author}{\bibfnamefont{R.}~\bibnamefont{Acciarri}}
  \bibnamefont{et~al.},
  \href{http://dx.doi.org/10.1088/1748-0221/5/05/p05003}{\bibinfo{journal}{Journal
  of Instrumentation} \textbf{\bibinfo{volume}{5}},
  \bibinfo{pages}{P05003}\bibinfo{year}{ (\bibinfo{year}{2010})}}.

\bibitem[{\citenamefont{Agnes et~al.}(2017{\natexlab{b}})}]{Agnes:2017grb}
\bibinfo{author}{\bibfnamefont{P.}~\bibnamefont{Agnes}} \bibnamefont{et~al.},
  \href{http://dx.doi.org/10.1088/1748-0221/12/10/P10015}{\bibinfo{journal}{Journal
  of Instrumentation} \textbf{\bibinfo{volume}{12}},
  \bibinfo{pages}{P10015}\bibinfo{year}{
  (\bibinfo{year}{2017}{\natexlab{b}})}}.

\bibitem[{\citenamefont{Agnes et~al.}(2021)}]{Agnes:2021zyq}
\bibinfo{author}{\bibfnamefont{P.}~\bibnamefont{Agnes}} \bibnamefont{et~al.},
  \href{http://dx.doi.org/10.1140/epjc/s10052-021-09801-6}{\bibinfo{journal}{Eur.
  Phys. J. C} \textbf{\bibinfo{volume}{81}},
  \bibinfo{pages}{1014}\bibinfo{year}{ (\bibinfo{year}{2021})}}.

\bibitem[{\citenamefont{Babicz et~al.}(2020)}]{Babicz_2020}
\bibinfo{author}{\bibfnamefont{M.}~\bibnamefont{Babicz}}  \bibnamefont{et~al.},
   \href{http://dx.doi.org/10.1088/1748-0221/15/09/p09009}{\bibinfo{journal}{Journal
  of Instrumentation} \textbf{\bibinfo{volume}{15}},
  \bibinfo{pages}{P09009}\bibinfo{year}{ (\bibinfo{year}{2020})}}.

\bibitem[{\citenamefont{Benson et~al.}(2018)\citenamefont{Benson, Orebi~Gann,
  and Gehman}}]{Benson:2017vbw}
\bibinfo{author}{\bibfnamefont{C.}~\bibnamefont{Benson}},
  \bibinfo{author}{\bibfnamefont{G.}~\bibnamefont{Orebi~Gann}},
  \bibnamefont{and} \bibinfo{author}{\bibfnamefont{V.}~\bibnamefont{Gehman}},
  \href{http://dx.doi.org/10.1140/s10052-018-5807-z}{\bibinfo{journal}{Eur.
  Phys. J. C} \textbf{\bibinfo{volume}{78}},
  \bibinfo{pages}{329}\bibinfo{year}{ (\bibinfo{year}{2018})}}.

\end{thebibliography}
\clearpage
\appendix


\end{document}